\documentclass[12pt]{iopart}
\usepackage{graphicx}% Include figure files
\usepackage{dcolumn,color}% Align table columns on decimal point
\usepackage[english]{babel}
\usepackage{bbm}
\usepackage{iopams}
\usepackage{url}

\expandafter\let\csname equation*\endcsname\relax
\expandafter\let\csname endequation*\endcsname\relax
\usepackage{amsmath}

\begin{document}
\title[Response Theory, Reaction Coordinates and Critical Phenomena]{Response Theory Identifies Reaction Coordinates and Explains Critical Phenomena in Noisy Interacting Systems}

\author{N Zagli $^1$ $^4$, V Lucarini $^2$ $^4$ and G A Pavliotis $^3$}

\address{$^1$ Nordita, Stockholm University and KTH Royal Institute of Technology - Hannes Alfvéns väg 12, SE-106 91 Stockholm, Sweden}
\address{$^2$ Department of Mathematics and Statistics, University of Reading - Reading, RG6 6AX, UK}
\address{$^3$ Department of Mathematics, Imperial College London - London, SW7 2AZ, UK}
\address{$^4$ Centre for the Mathematics of Planet Earth, University of Reading - Reading, RG6 6AX, UK}
\ead{niccolo.zagli@su.se}

\vspace{10pt}

\begin{abstract}
We consider a class of nonequilibrium systems of interacting agents with pairwise interactions and quenched disorder in the dynamics featuring, in the thermodynamic limit,  phase transitions. 
%Firstly, 
We provide conditions on the microscopic structure of interactions among the agents that lead to a dimension reduction of the system in terms of a finite number of reaction coordinates.
%Secondly, We show that such
 Such reaction coordinates prove to be proper nonequilibrium thermodynamic variables as they carry information on correlation, memory and resilience properties of the system. Phase transitions can be identified and quantitatively characterised as singularities of the complex valued susceptibility functions associated to the reaction coordinates. We provide analytical and numerical evidence of how the singularities affect the physical properties of finite size systems.

\end{abstract}

\vspace{2pc}

%\submitto{\jpa}
\maketitle
\section{Introduction}
Interacting agent models are at the basis of the microscopic description of the rich variety of collective emergent phenomena that high dimensional complex systems often exhibit. Common applications of such models range from synchronisation of nonlinear oscillators ~\cite{SakaguchiRossler,Pikovsky2003,Pecora1998,Pecora2015,Eroglu2017,OTT200229,Kuramoto,PolitiClusellaBiHarmonic}, see also the recent special issue \cite{PikovskySpecialIssue2023}, phase transitions in complex energy landscapes \cite{Gomes,GomesPavliotis2017} to opinion dynamics and consensus formation \cite{HasgelmannKrause,Goddard2022}, socio-economic siences \cite{NaldiParentiToscani,toscani2014}, life sciences \cite{DaiPra}, the dynamics of the brain \cite{MontbrioPazRoxin2015}, formation of swarms \cite{Carrillo2014,Carrillo2010reviewSwarming}, dynamical networks \cite{Carrillo:2020aa}, self-gravitating systems \cite{Chavanis2014,SelfGravitating} and algorithms for optimisation and training of neural networks \cite{borovykh2020stochastic,reich2020,RotskoffVandenEijnden2022}.
In the thermodynamic limit, such systems can exhibit phase transitions resulting from the interplay between the interaction among the agents, their internal dynamics and the noise. Such critical phenomena are associated with the so called critical slowing down phenomenon, that is by long lasting, persistent fluctuations exhibiting statistics far away from Gaussianity, see  \cite{Dawson} for a detailed description of the critical fluctuations for an equilibrium continuous phase transition. 
\subsection{From a microscopic to a macroscopic description of the system}
From an operational point of view, the investigation of these critical phenomena heavily relies on the identification of a suitable set of observables, commonly referred to as order parameters (OPs), collective variables (CVs) or reaction coordinates (RCs),  acting as thermodynamic variables by providing a low dimensional description of the macroscopic feature of the system. Whereas it is true that all the three notions of OP, CV and RC are related to a dimension reduction procedure provided by a projection from a high dimensional system to a reduced state space, they refer to different features of the system.  Along the lines of \cite{Rogal2021}, we define a CV as any function of the full phase space. An OP is then a suitable combination of CVs that is able to distinguish between macroscopic, possibly metastable, states of the sytem. As  some relevant parameters of the system are varied, one wishes to identify critical points by observing asymptotic properties of the OPs, that is by constructing phase diagrams. On the other hand, the term RC refers to a suitable OP that is able to capture time dependent properties of the system, e.g. relaxation phenomena or transitions between metastable states.
\\
A major challenge in the identification of critical phenomena is that, while order parameters or reaction coordinates can in many cases be deduced for equilibrium systems using, \textit{e.g.}, symmetry arguments, the identification of reaction coordinates in nonequilibrium settings is far less trivial \cite{Ma2005,Laio2006,Rogal2021} and is tailored to the specific problem at hand. 
Here we provide a constructive, as opposed to ad hoc, criterion to identify a set of reaction coordinates $\{ A_i \}$ for a large class of interacting agent systems based on the microscopic interaction structure among the agents. Inspired by the success of response theory in explaining critical phenomena in a class of interacting systems \cite{FirstPaper,ZagliLucariniPavliotis,ZagliPavliotisLucarini2023,Topaj2001}, we show that, according to this perspective, the $\{ A_i \}$ not only represent OPs for the system but correspond to suitable RCs as they act as proper nonequilibrium thermodynamic variables. On the one hand, their stationary asymptotic properties identify the different macroscopic states of the system. On the other hand, they also carry information on correlation properties and resilience of the system to perturbations. Most importantly,  we show that such observables represent the optimal observables to fully characterise the development of a critical phenomenon as the critical, strongly peaked, resonances in their linear response to external perturbations allow for a quantitative description of the critical behaviour. 
\subsection{Linear Response}
Linear Response Theory provides a powerful framework to predict how statistical properties of a system change as a result of weak, yet arbitrary exogenous perturbations. Originally developed for statistical mechanical systems near equilibrium,  the theory has much larger validity \cite{ruelle_review_2009} with applications ranging from driven, strongly dissipative systems such as the climate \cite{GhilLucarini2020,LucariniChekroun2023} to neural networks \cite{Cessac2019,Cessac2021,SoonHoeLim2021} and financial markets \cite{Puertas2021}. In brief, a linear response perspective corresponds to the following conceptual and mathematical framework. Given a system with a unique, physical, ergodic (not necessarily smooth) measure $\mu_0$, linear response theory aims at evaluating changes in the expectation value
\begin{equation}
   \langle \Phi \rangle_0 := \int \Phi(\mathbf{x}) \mu_0(\mathrm{d}\mathbf{x}) = \lim_{T \rightarrow + \infty} \frac{1}{T} \int_0^T \Phi(\mathbf{x}(t)) \mathrm{d}t 
\end{equation}
where $\Phi$ is a generic observable of the system, $\mathbf{x}(t)$ is a typical trajectory and the last equality hold because of ergodicity. Linear response theory is valid for a vast class of systems ranging from deterministic dynamical systems \cite{ruelle_review_2009,liverani2006}, Markov chain models \cite{SantosJSP} to continuous stochastic systems  \cite{Marconi2008,pavliotisbook2014} where it can be justified on rigorous mathematical grounds \cite{HairerMajda2010}. We here consider a finite dimensional system described by a set of Ito stochastic differential equations with deterministic drift $\mathbf{\mathcal{F}} : \mathbb{R}^d \to \mathbb{R}^d$ and noise law given by the matrix valued function $\boldsymbol{\Sigma}: \mathbb{R}^d \to \mathbb{R}^{d \times q}$ 
\begin{equation}
\label{eq: finite dimensional unperturbed}
    \mathrm{d}\mathbf{y}(t) = \mathbf{\mathcal{F}}(\mathbf{y})\mathrm{d}t + \boldsymbol{\Sigma}(\mathbf{y})\mathrm{d}\mathbf{W}_t
\end{equation}
where $\mathrm{d}\mathbf{W}_t$ is a $q-$dimensional Wiener process. If we assume that the stochastic process is hypoelliptic \cite{pavliotisbook2014}, the invariant measure $\mu_0$ associated to the above stochastic process is absolutely continuous with respect to the Lebesgue measure with smooth probability density $\rho_0$, that is $\mu_0(\mathrm{d}\mathbf{y}) = \rho_0(\mathbf{y})\mathrm{d}\mathbf{y}$, even when the deterministic drift $\mathbf{\mathcal{F}}$ induces a chaotic dynamics. We perturb the system by letting $\mathbf{\mathcal{F}} \to \mathbf{\mathcal{F}} + \varepsilon \mathbf{Y}(\mathbf{y}) T(t)$ where the bounded function $T(t)$ represents the time modulation of the external perturbation and $Y(y)$ its phase space state dependence. In this setting, the statistical properties of the system change as  
\begin{equation}
\langle \Phi \rangle(t) = \langle \Phi \rangle_0 + \varepsilon \langle \Phi \rangle_1(t) + o(\varepsilon), \quad \text{with} \quad \langle \Phi \rangle_1(t) = \left( G_\Phi \star T \right)(t)  
\end{equation}
where $G_\Phi$ is the Green's function associated to the observable $\Phi$. The Green's function $G_\Phi$ encodes all the positive and negative feedbacks of the system and allows to evaluate the response of the system to general time dependent forcings. Response formulas provide a useful decomposition of any Green's function as\footnote{In expression \eqref{eq: decomposition of Green's function finite dimensional} the contribution of the continuous part of the spectrum is neglected, which is equivalent to assuming that the Markov semigroup describing the dynamics is quasi compact \cite{EngelNagel2000}.  }  
\begin{equation}
\label{eq: decomposition of Green's function finite dimensional}
    G_\Phi(t) = \sum_{k=1} \sum_{l=1}^{m_k} g^{(l)}_k t^{l-1} e^{\lambda_k t} 
\end{equation}
where the complex valued exponential rates $\lambda_k \in \mathbb{C}$ and $\mathbf{Re}\lambda_k < 0$ are the eigenvalues of the generator of the Markov semigroup associated to the unperturbed dynamics \eqref{eq: finite dimensional unperturbed} and are referred to as stochastic Ruelle Pollicott resonances  and $g_k^{(l)}$ are suitable coefficients \cite{GutierrezLucarini2022,chekroun2019c,lasota}.  Moreover, if the eigenvalue $\lambda_k$ is not simple, that is its multiplicity is $m_k >1$, modulating polynomial terms arise in the expression for the Green's function. 
We remark that the  exponential rates $\lambda_k$ represent an intrinsic property of the unperturbed dynamics and correspond to natural relaxation and oscillatory timescales of the system. Green's functions are thus the relevant quantities to be investigated in order to detect the onset of critical instabilities of general systems. In the context of interacting systems, it has been shown that long lasting and persistent modes with exponential rates of small real part are associated to poorly damped instabilities due to phase transitions in the thermodynamic limit \cite{ZagliLucariniPavliotis}. 
\\
It is important to stress that the choice of the observable under investigation remains a critical issue in detecting criticalities as the coefficients $g^{(l)}_k$ are not an intrinsic property of the system but depend both on the observable $\Phi$ and the state dependence $\mathbf{Y}(\mathbf{y})$ of the perturbation  \cite{GutierrezLucarini2022}. In particular, if the observable $\Phi$ has vanishingly small projection on the most unstable mode, its associated Green's function $G_\Phi$ would not represent a good proxy for the detection of the critical behaviour of the system.
\subsection{This paper}
The main contributions of this paper can be summarised as follows:
\begin{itemize}
    \item We provide a constructive way to define a finite set of RCs $\{ A_i \}$ for the thermodynamic limit of an ensemble of interacting agents. The RCs correspond to an effective dimension reduction of the system in terms of a few relevant observables. We provide mathematical conditions on the interaction structure among the agents for which the construction of RCs is exact. In the context of synchronisation phenomena for ensembles of phase oscillators with purely harmonic coupling, dimension reduction is usually obtained by invoking the Ott-Antonsen ansatz and generalisations thereof \cite{OttAntonsen,TyulkinaPRL2018,Goldobodin2021,CestnikPikovsky2022}.  Our results are instead valid for any coupling structure and even in settings where phase reduction techniques of nonlinear oscillators \cite{Wilson2019,Nakao} cannot be applied. In this context, this paper represents a generalisation of \cite{OTT200229} to settings with (a) possibly multiplicative, white noise sources and (b) an interaction structure that is not a priori assumed to depend on (functions of) the empirical mean of the ensemble.
    \item  We derive response formulas for the thermodynamic limit of the ensemble of agents. We show that the response of general mean field observables of the system is mediated by these RCs. In particular, we show that the coupling among the agents manifest itself in the thermodynamic limit as a memory contribution, conveyed by the RCs, to the sensitivity properties of the system. We prove that the macroscopic Green's function describing the response of the system at a macroscopic level encodes feedbacks stemming from both microscopic properties of each agent and the interaction structure among them.  
    \item We show that the investigation of  criticalities arising from phase transitions are associated, in the thermodynamic limit, to the divergent behaviour of Green's functions relative to the RCs. Phase transitions can be quantitatively identified as the singular part of the complex valued susceptibility, the Fourier Transform of the Green's Function. These singularities can be characterised by a pole-residue pair $(\omega^*,\kappa)$ with $\omega^* \in \mathbb{R}$ and $\kappa \in \mathbb{C}$. The pole $\omega^*$ corresponds to frequencies of the driving force that would critically destabilise the system, whereas the residue $\kappa$ determines the strength and detailed nature of the singularity.
    \item We provide analytical and numerical evidence that the analysis of response properties of the RCs in a finite size ensemble of agents is able to capture the signature of a phase transition as a resonance around the singular thermodynamic behaviour. We show that both the pole $\omega^*$ and the residue $\kappa$ can be inferred in a finite system. In particular, we illustrate the physical role that the residue $\kappa$ plays on response properties of the system. 
\end{itemize}
The paper is structured as follows. In section \ref{sec: The model} we introduce the class of interacting systems under investigation and we describe when a dimension reduction in terms of RCs is possible. Section \ref{sec: Response Formulas} is dedicated to the description of linear response properties of the system in terms of the RCs. In section \ref{sec: Kuramoto model} we investigate the critical response properties of a finite ensemble of Kuramoto oscillators. In the conclusions we mention future research directions of this linear response framework to investigate resilience features of general interacting systems.
\section{The interacting agent models}
\label{sec: The model}
We consider an ensemble of $N$  interacting $M$-dimensional systems $\{ \mathbf{x}^k \}_{k=1}^N$ whose dynamics is described by the following stochastic differential equations
\begin{equation}
\label{eq: N particle system quenched}
    \mathrm{d}\mathbf{x}^k = \mathbf{\mathbf{F}}_\alpha(\mathbf{x}^k; \mathbf{h}^k) \mathrm{d}t - \frac{\theta}{N}\sum_{j=1}^N  \mathbf{K}(\mathbf{x}^k,\mathbf{x}^j) \mathrm{d}t+ \sigma \mathbf{s}(\mathbf{x^k})\mathrm{d}\mathbf{W}^{(k)} 
\end{equation}
The smooth vector field $\mathbf{F}_\alpha(\mathbf{x}; \cdot) : \mathbb{R}^M \to \mathbb{R}^M$, possibly depending on a set of parameters $\alpha$, defines the local dynamics of each agent corresponding to, in general, nonequilibrium,  dissipative conditions.The (pair-wise) interactions among the systems are introduced via the function $\mathbf{K} : \mathbb{R}^M \times \mathbb{R}^M \to \mathbb{R}^M$ with $\theta$ representing the coupling strength. The function $\mathbf{K}$ is assumed to be antisymmetric in its arguments, $\mathbf{K}(\mathbf{x},\mathbf{y}) = -\mathbf{K}(\mathbf{y},\mathbf{x})$, to model Newton's third law of motion for the internal forces acting between the systems. The volatility matrix $\mathbf{s}: \mathbb{R}^M \to \mathbb{R}^{M \times M}$ determines the state dependent noise term, with $\sigma$ being its amplitude and $\mathrm{d}\mathbf{W} = (\mathrm{d}\mathbf{W}^{(1)}, \dots, \mathrm{d}\mathbf{W}^{(N)})$ representing a $M \times N$ Brownian motion. In  the following we will adopt the Itô's convention. Furthermore, we introduce a source of quenched disorder by letting the function $\mathbf{F}_\alpha$ depend on a vector of parameters $\mathbf{h}^k \in \mathbb{R}^m$ drawn from a fixed known distribution $\mu$, that is $\mathbf{h}^k \sim \mu$ $\forall k = 1, \dots, N$. The microscopic architecture of the system given by $\mu$ can be interpreted either as an intrinsic property of the system, such as the natural frequencies of an ensemble of oscillators, or as a model error feature arising from partial knowledge of the microscopic properties of the agents.
\\
In the thermodynamic limit $N \to + \infty$, the mean field nature of the coupling allows to obtain a macroscopic hydrodynamic description of the ensemble of agents  \cite{McKean2,Snitz,ChaintronDiez2022}. More specifically, we define the empirical density of the $N$ particle system as $\rho_N(\mathbf{x},\mathbf{h},t) = \frac{1}{N}\sum_{j=1}^N \delta(\mathbf{x} - \mathbf{x}^j)\delta(\mathbf{h} - \mathbf{h}^j)$, where $\delta(\cdot)$ represents the Dirac delta distribution. Under general conditions, it is possible to show that the system exhibits propagation of molecular chaos \cite{DaiPra,PraHollander}, i.e. , fixed $\mathrm{T} \in \mathbb{R}$ and given a chaotic initial condition $\mathbf{x}^k(t=0) \sim \hat{\rho}(\mathbf{x})$ $\forall k=1,\dots,N$ , the empirical density $\rho_N$ converges weakly to $\rho(\mathbf{x},t;\mathbf{h})\mu(\mathbf{h})$ where the the one-particle distribution $\rho(\mathbf{x},t;\mathbf{h})$ satisfies the nonlinear Fokker Planck Equation (NFPE) 
\begin{equation}
\begin{split}
\label{eq: General Equation quenched quenched}
\partial_t \rho &= - \nabla \cdot \left[\left(\mathbf{\mathbf{F}}_\alpha(\mathbf{x},\mathbf{h}) - \theta \int \left(\mathbf{K}\star_2 \rho\right) \mu(\mathbf{h})\mathrm{d}\mathbf{h} \right) \rho \right] + \\
&+ \frac{\sigma^2}{2} \nabla^2 : \left( \mathbf{s}(\mathbf{x})\mathbf{s}^T(\mathbf{x}) \rho \right) 
\end{split}
\end{equation}
where $\rho(\mathbf{x},0;\cdot)  =\hat{\rho}(\mathbf{x})$ and $t \in [0,\mathrm{T}]$. In equation \eqref{eq: General Equation quenched quenched} the drift term depends on the probability distribution $\rho$ itself through $(\mathbf{K} \star_2 \rho)(\mathbf{x};\mathbf{h}) = \int \mathbf{K}(\mathbf{x},\mathbf{y}) \rho(\mathbf{y},t;\mathbf{h}) \mathrm{d}\mathbf{y}$ originating from the coupling among the agents and $\nabla^2 : \left( \mathbf{s}\mathbf{s}^T \rho \right)  = \sum_{i,j,k=1}^M\partial^2_{ij}\left( s_{ik}(\mathbf{x})s_{jk}(\mathbf{x})\rho  \right)$ represents the state dependent diffusive part. We remark that the microscopic architecture of the system manifests itself in equation \eqref{eq: General Equation quenched quenched} as an expectation value over the distribution $\mu$. In particular, the homogeneous case where no quenched disorder is present is obtained by setting $\mu(\mathbf{h}) = \delta(\mathbf{h}-\mathbf{h}_0)$ where $\mathbf{h}_0$ is a constant parameter. We observe that equation \eqref{eq: NLFPEquenched} represent a nonlinear and nonlocal (integro-differential) equation and can thus support multiple coexisting stationary measures, corresponding to different macroscopic state of the system. Furthermore, exchange of stability of such measures, as the parameters of the system are varied, correspond to phase transition for the system \cite{FrankBook,Carrillo:2020aa}. 
\\
We remark that the phenomenon of propagation of molecular chaos stated above is equivalent to a mean field approximation of the Fokker Planck associated to \eqref{eq: N particle system quenched}, where the N-particle probability distribution is assumed to be factorised as a product of one-particle distributions $\rho(\mathbf{x},t;\mathbf{h})$. This corresponds to the simplest closure scheme, characterised by considering all agents as statistically independent, of the BBGKY hierarchy \cite{MartzelAslangul2001}, an infinite set of equations describing the correlations among the N subsystems. It is known that, at a phase transition point, correlations can no longer  be neglected at a macroscopic level \cite{Dawson}. Thus, near a phase transition point the NFPE is not expected to capture the behaviour of fluctuations, correlation and relaxation properties of macroscopic observables. We mention that the Linear Response perspective described later in section \ref{sec: Response Formulas} and \ref{sec: Kuramoto model} can be interpreted as an approach to characterising critical fluctuations.
\subsection{Identification of Reaction Coordinates} 
\noindent
Equation \eqref{eq: General Equation quenched quenched} represents a coarse grained, mesoscopic, description of the system valid for a big, strictly infinite, ensemble of agents. However, even at this coarse grained level, this equation represents an infinite dimensional problem as the drift coefficient depends on the full one-particle probability distribution through the convolution product $(\mathbf{K} \star_2 \rho$. It is then fundamental to find conditions on the dynamics of the system for which suitable OPs or RCs can be found.
It turns out that the characterisation of RCs for these systems is strictly related to the properties of the interaction kernel $\mathbf{K}$. We consider a separable interaction kernel, that is, admitting a decomposition
\begin{equation}
\label{eq: separable kernel}
K_i(\mathbf{x},\mathbf{y}) = \sum_{l = 1}^{d} k_{i;l}(\mathbf{x}) a_{i;l}(\mathbf{y})
\end{equation}where $d \in \mathbb{N}$, $d < + \infty$. We remark that it is always possible to assume that the functions in the above decomposition are linearly independent \cite{Kress2014lie}. In this case, the effect of the nonlinearity in equation  \eqref{eq: General Equation quenched quenched} can be written in terms of a finite number of suitable observables rather than on the full probability distribution $\rho$ since, using  \eqref{eq: separable kernel}, the interaction term results in

\begin{equation}
\begin{split}
    \label{eq: sep kernel2}
\int \left(K_i \star_2 \rho\right)(\mathbf{x};\mathbf{h}) \mu(\mathbf{h})\mathrm{d}\mathbf{h} &= \int \int  K_i(\mathbf{x},\mathbf{y}) \rho(\mathbf{y},t;\mathbf{h}) \mu(\mathbf{h})\mathrm{d}\mathbf{h}\mathrm{d}\mathbf{y} = \\
&= \sum_{l=1}^d k_{i;l}(\mathbf{x}) \langle \langle a_{i;l} \rangle \rangle(t) := \mathcal{K}_i(\mathbf{x};\{\langle \langle a_{i;l} \rangle \rangle \}_l)
\end{split}
\end{equation}
where $\langle \langle \cdot \rangle \rangle$ represents the double expectation value with respect to $\rho(\mathbf{x},t; \mathbf{h})$ and $\mu$. Equation \eqref{eq: General Equation quenched quenched} can now be written as  
\begin{equation}
\begin{split}
\label{eq: NLFPEquenched}
    \partial_t \rho &= - \nabla \cdot \bigg( \big(\mathbf{\mathbf{F}}_\alpha(\mathbf{x},\mathbf{h}) - \theta \boldsymbol{\mathcal{K}}\left(\mathbf{x}, \langle \langle \mathbf{A} \rangle \rangle\right)  \big) \rho\bigg) + \\
&+ \frac{\sigma^2}{2} \nabla^2 : \left( \mathbf{s}(\mathbf{x})\mathbf{s}^T(\mathbf{x}) \rho \right) 
     := \mathcal{L}_{\langle\langle \mathbf{A} \rangle\rangle} \rho
\end{split}
\end{equation}
where $\langle \langle \mathbf{A}(\mathbf{x})\rangle \rangle = \{ \langle \langle a_{i;l}(\mathbf{x})\rangle \rangle \}^{M,d}_{i=1,l=1} \in \mathbb{R}^p$ is a finite vector ($p$ not necessarily equals to $d$ and $p \ll N$, see later discussion) of observables  and $A_j(\mathbf{x})$ $j=1,\dots,p$ are linearly independent functions. We have also defined the nonlinear operator $\mathcal{L}_{ \langle \langle \mathbf{A} \rangle \rangle }$ depending on the probability distribution through the vector $\langle \langle \mathbf{A} \rangle \rangle$.  Following the terminology of \cite{FrankBook}, this scenario corresponds to a finite nonlinearity dimension equals to $p$ for equation \eqref{eq: General Equation quenched quenched} and it corresponds to a natural dimension reduction of the ensemble of agents in terms of the variables $\langle \langle \mathbf{A} \rangle \rangle $. 
\\
Separable kernels $\mathbf{K}(\mathbf{x},\mathbf{y})= \nabla \mathcal{U}(\mathbf{x}-\mathbf{y})$ arising from interaction potentials are routinely used in applications, with radial potentials of the form $\mathcal{U}(\mathbf{x})= u(|\mathbf{x}|)$ being a common choice. A paradigmatic example is represented by quadratic interactions  $u(x) = \frac{x^2}{2}$, yielding a separable kernel $\mathbf{K}(\mathbf{x},\mathbf{y}) = \mathbf{x} - \mathbf{y}$. This corresponds to linear forces among the particles trying to synchronise them towards their centre of mass $\bar x = \frac{1}{N} \sum_j{\mathbf{x}}_j(t)$ and has numerous applications, see our previous work \cite{FirstPaper,ZagliLucariniPavliotis} on this. In this case, $\forall i=1,\dots,M$, the expansion \eqref{eq: separable kernel} of the interaction kernel holds with $d=2$ and  $(f_{i;1},g_{i;1},f_{i;2},g_{i;2})=(x,0,0,-y)$  resulting in a non linear dimension $p=
M$ for equation \eqref{eq: NLFPEquenched} given by the observables $\langle \langle \mathbf{A}(\mathbf{x}) \rangle \rangle = \langle \langle \mathbf{x} \rangle \rangle$. Similarly, higher order polynomials $u(x) = \sum_{k=2}^n u_k x^k$ with $n \geq 2$ give rise to separable kernels and a finite nonlinearity dimension. As a further example, we mention separable kernels provided by trigonometric interactions of any order with applications to synchronisation phenomena, see \cite{Battle1977,PolitiClusellaBiHarmonic} and the section below on the Kuramoto model. We refer the reader to the conclusions section for a discussion regarding non separable kernels. 
\\ 
We remark that equation \eqref{eq: NLFPEquenched} can be interpreted as a thermodynamic formulation of the system with the $\langle \langle \mathbf{A} \rangle \rangle$s being a generalisation of thermodynamic variables to nonequilibrium settings \cite{Zubarev1996}. Firstly, stationary distributions $\rho_0 = \rho_0(\mathbf{x};\mathbf{h})$ of \eqref{eq: NLFPEquenched} can be parametrised in terms of the stationary values of the thermodynamic variables $\langle \langle \mathbf{A} \rangle \rangle_0$ according to the self consistency equation given by the stationary condition $\mathcal{L}_{0} \rho_0 =0$ where $\mathcal{L}_0= \mathcal{L}_{\langle \langle \mathbf{A}\rangle\rangle_0}$ is the operator defined in \eqref{eq: NLFPEquenched} evaluated at stationarity. 
This is better understood in the case of homogeneous equilibrium systems, characterised by a gradient dynamics $\mathbf{F}_{\alpha}(\mathbf{x})= - \nabla V_\alpha(\mathbf{x})$, an interaction potential $\mathcal{U}(\mathbf{x})$, thermal noise $\mathbf{s}(\mathbf{x}) = \mathbf{1}_{M\times M}$ and no quenched disorder. In such systems, stationary solutions of \eqref{eq: General Equation quenched quenched} satisfy the self consistency equation \cite{Carrillo:2020aa} 
\begin{equation}
    \rho_{eq}(\mathbf{x}) = \frac{1}{Z}\exp\left(-\frac{2}{\sigma^2}\left( V_\alpha(\mathbf{x}) - \mathcal{U}\star \rho_{eq} \right) \right)
\end{equation}
where $Z$ is a normalisation constant. Assuming a separable kernel, equilibrium stationary solutions are parametrised by a finite number of observables $\langle \mathbf{A} \rangle_{eq}$ since they can be written as $\rho_{eq}\left(\mathbf{x};\langle \mathbf{A} \rangle_{eq} \rangle\right) = \frac{1}{Z}\exp\left(-\frac{2}{\sigma^2}\left( V_\alpha(\mathbf{x}) - f(\mathbf{x},\langle \mathbf{A} \rangle_{eq}) \right)\right)$ where $f$ is a suitable function deriving by the convolution product $\mathcal{U}\star \rho_{eq}$, similarly to equation \eqref{eq: sep kernel2}.
\\Moreover, the $ \langle \langle \mathbf{A}(\mathbf{x}) \rangle \rangle$s represent proper RCs of the system as they govern time dependent properties of the ensemble of agents, such as its response, correlation and memory features, as we elucidate in the next section.
\section{Response Formulas}
\label{sec: Response Formulas}
We remark that a stationary state $\rho_0 $ of equation \eqref{eq: NLFPEquenched} is characterised by the constant set of observables $\langle \langle \mathbf{A} \rangle \rangle_0$, where the subscript denotes that the expectation value is taken with respect to $\rho_0$ (and $\mu$).
We perturb such state with a time and state dependent perturbation by letting $\mathbf{F}_{\alpha}(\mathbf{x};\mathbf{h}) \rightarrow \mathbf{F}_\alpha(\mathbf{x};\mathbf{h}) + \varepsilon T(t)\mathbf{X}(\mathbf{x})$ and we write $\rho(\mathbf{x},t;\mathbf{h}) \approx \rho_0(\mathbf{x};\mathbf{h}) + \varepsilon \rho_1(\mathbf{x},t;\mathbf{h})$. We observe that, since $\langle \langle \mathbf{A} \rangle \rangle \approx \langle \langle  \mathbf{A} \rangle \rangle_0 + \varepsilon \langle \langle \mathbf{A} \rangle \rangle_1(t)$, where the last subscript denotes the expectation value with respect to $\rho_1$, the perturbation to the nonlinear part of the drift term can be written as
\begin{equation}
\boldsymbol{\mathcal{K}}\left(\mathbf{x}, \langle \langle\mathbf{A} \rangle \rangle\right) \approx \boldsymbol{\mathcal{K}}\left(\mathbf{x}, \langle\langle  \mathbf{A} \rangle\rangle_0 \right) + \varepsilon \mathbf{J}\left(\mathbf{x} \right)\langle\langle \mathbf{A} \rangle\rangle_1(t)
\end{equation} where the matrix $\mathbf{J} \in \mathbb{R}^{M \times p}$ provides the information on how the drift term changes, in a linear approximation, with respect to the observables $\langle \langle \mathbf{A} \rangle \rangle$ and is defined as
\begin{equation}
\label{eq: matrix J}
J_{ij}\left( \mathbf{x} \right) = J_{ij}(\mathbf{x},\langle\langle \mathbf{A} \rangle\rangle_0) = \frac{\partial \mathcal{K}_i}{\partial \langle\langle A_j \rangle\rangle}\left( \mathbf{x}, \langle\langle \mathbf{A} \rangle\rangle_0 \right)
\end{equation}
From \eqref{eq: NLFPEquenched} we obtain an equation for the perturbation to the stationary measure
\begin{equation}
\begin{split}
    \label{eq: correction to invariant measure}
    \rho_1(&\mathbf{x},t;\mathbf{h}) = \int_{0}^t T(s) e^{\left(t-s\right)\mathcal{L}_{0}  }\mathcal{L}_p \rho_0 \mathrm{d}s +\\
    &+\theta \sum_{j=1}^{p} \int_{0}^t \langle \langle A_j \rangle\rangle_1(s) \sum_{k=1}^M e^{\left( t-s\right) \mathcal{L}_{0}} \partial_{x_k} \left( J_{kj}\left(\mathbf{x} \right)\rho_0\right) \mathrm{d}s
\end{split}
\end{equation}
where $\mathcal{L}_p \cdot := - \nabla \cdot \left( \mathbf{X}(\mathbf{x}) \quad \!\!\!\! \cdot \quad \!\!\!\!  \right)$ is the perturbation operator.
%$\mathcal{L}_0=\mathcal{L}_{\langle \langle \mathbf{A}\rangle\rangle_0}$ is the operator defined in \eqref{eq: NLFPEquenched} evaluated at the stationary state
We now evaluate the perturbation to the observable $A_i(\mathbf{x})$ by taking the expected value with respect to $\rho_1$ and $\mu$ to obtain
\begin{equation}
\begin{split}
\label{eq: expected value 1 quenched}
\langle \langle A_i \rangle \rangle_1(t) = \left( G_i \star T \right)(t) + \theta \sum_{j=1}^p \left(  Y_{ij} \star \langle \langle A_j \rangle \rangle_1  \right)(t)
%&= \int_{-\infty}^{+\infty} T(s) G_i(t-s) \mathrm{d}s + \\
%&+ \theta \sum_{j=1}^p\int_{-\infty}^{+\infty} \langle \langle A_j \rangle \rangle(s) Y_{ij}(t-s) \mathrm{d}s
\end{split}
\end{equation}
where $\star$ denotes the convolution product and we have defined the microscopic Green's functions %$G_i(t) = \int G_i(t;\mathbf{h})\mu(\mathbf{h})\mathrm{d}\mathbf{h}$ and $Y_{ij}(t)  = \int Y_{ij}(t;\mathbf{h})\mu(\mathbf{h})\mathrm{d}\mathbf{h} $ obtained from
\begin{subequations}
\begin{align}
\label{eq: microscopic Green Function G}
G_i(t) &=  \Theta(t)\int \mathrm{d}\mathbf{h}\mu(\mathbf{h})\int \mathrm{d}\mathbf{x} A_i(\mathbf{x}) e^{t\mathcal{L}_{0} } \mathcal{L}_p \rho_0,
\\
\label{eq: microscopic Green Function Y}
Y_{ij}(t) &= \Theta(t)\int \mathrm{d}\mathbf{h}\mu(\mathbf{h}) \int \mathrm{d}\mathbf{x} A_i(\mathbf{x}) e^{t\mathcal{L}_{_0} } \sum_{k=1}^M \partial_{x_k} \left( J_{kj}\left(\mathbf{x}\right) \rho_0\right). 
\end{align}
\end{subequations}
We observe that $G_i$ depends on the perturbation whereas $Y_{ij}$ does not, representing thus an intrinsic property of the system describing its internal feedbacks, see \cite{FirstPaper}.  
The coupling among the agents manifests itself as a memory term in \eqref{eq: expected value 1 quenched} conveyed by the variables $\langle \langle A_j\rangle \rangle$ through the microscopic Green's Function $Y_{ij}$. 
Taking the Fourier transform of \eqref{eq: expected value 1 quenched} we obtain
\begin{equation}
\label{eq: response in frequency}
\sum_{j=1}^p P_{ij}(\omega) \langle \langle A_j \rangle \rangle_1(\omega) = T(\omega) \chi_{i}(\omega)
\end{equation}where we have defined the $p \times p$ matrix 
\begin{equation}
\label{eq: renormalisation matrix}
P_{ij}(\omega) = \delta_{ij} - \theta Y_{ij}(\omega)
\end{equation}and $\chi_i(\omega)$  ($Y_{ij}(\omega)$) are the Fourier transform of the microscopic Green functions $G_i(t)$ ($Y_{ij}(t)$) respectively. Equations \eqref{eq: response in frequency} and \eqref{eq: renormalisation matrix} completely characterise the response of statistical properties of the system to perturbations in terms of the observables $\langle \langle \mathbf{A} \rangle \rangle$. When $\chi_i(\omega)$ is an analytic function in the upper complex $\omega$ plane and $P_{ij}(\omega)$ is invertible the response of the system is smooth and given by
\begin{equation}
\label{eq: response in frequency with macroscopic susceptibility}
    \langle \langle \mathbf{A} \rangle \rangle (\omega) = \tilde{\boldsymbol{\chi}}(\omega) T(\omega) , \quad \tilde{\boldsymbol{\chi}}(\omega) = \mathbf{P}^{-1}(\omega)\boldsymbol{\chi}(\omega)
\end{equation}
where $\tilde{\boldsymbol{\chi}}$  is the complex valued macroscopic susceptibility associated to the RCs. In this case, $\tilde{G}_i(t)$, the inverse Fourier transform of $\tilde{\chi}_i(\omega)$, defines a causal macroscopic Green's function describing the response properties of the system as 
\begin{equation}
    \langle \langle A_i(t) \rangle\rangle_1(t) = (\tilde{G}_i \star T)(t).
\end{equation}
From an operational point of view, the macroscopic Green's function $\tilde{G}_i(t)$ is the quantity that is experimentally accessible at a macroscopic level. This function encodes all positive and negative feedbacks of the system, including feedbacks stemming from microscopic single agent features and from the interaction among them. We remark that the macroscopic Green's function $\tilde{G}_i(t)$, being the inverse Fourier transform of the macroscopic susceptibility defined in \eqref{eq: response in frequency with macroscopic susceptibility}, has a non trivial expression in terms of the microscopic Green's function $G_i(t)$ and $Y_{ij}(t)$. 
\\
Criticalities in the response of the system arise when the negative feedbacks are no longer successful in damping the external perturbations and are signalled by the fact that some modes of the macroscopic Green's function $\tilde{G}_i(t)$ do not decay in time. More specifically, the response of the system breaks down as a result of a pole in $\tilde{\boldsymbol{\chi}}(\omega)$ when either $\chi_i(\omega)$ develops a singularity, corresponding to a destabilisation of each agent in the system \cite{FirstPaper}, or when $P_{ij}(\omega)$ becomes noninvertible. Equation \eqref{eq: renormalisation matrix} shows that the latter case arises due to endogenous instabilities associated with positive feedbacks resulting from the coupling among the agents. This identifies the occurrence of a phase transition in the system as it is intimately related to taking the thermodynamic limit \cite{FirstPaper}. We remark that, generally, a linear stability analysis of nonlinear Fokker Planck equations such as \eqref{eq: NLFPEquenched}, e.g. for the Kuramoto model described below, involves the investigation of spectral properties of (infinite dimensional) linear integral operators \cite{FrankBook,AcebronBonilla2005}. However, the identification of RCs allows to simplify the stability problem into the analysis of the finite dimensional matrix $P_{ij} \in \mathbb{C}^{p \times p}$ encoding their intrinsic response properties.   
\section{Critical phenomena for the Kuramoto model}
\label{sec: Kuramoto model}
We elucidate the above results by investigating critical phenomena in the Kuramoto model \cite{AcebronBonilla2005,GuptaCampaRuffo2014}, the paradigmatic example of synchronisation phenomena of phase oscillators. We remark that, even if the Kuramoto model has been thoroughly investigated, a stability theory of the stationary solutions of the Kuramoto model is still lacking, see \cite{IatsenkoetalPRL2013} for partial results in this direction in deterministic settings. The goal of this section is to  shed light on the physical mechanisms underlying the onset of synchronisation encoded in the properties of the Green's function of the system. In particular, we are interested in showing how the critical response of a finite dimensional system relates to the properties of the complex valued susceptibility $\tilde{\boldsymbol{\chi}}(\omega)$.
\\
The Kuramoto model is defined by the following SDEs
\begin{equation}
\label{eq : Kuramoto model}
    \mathrm{d}x_k = \big[ h_k -\frac{\theta}{N} \sum_{j=1}^N \sin\left( x_k - x_j \right) \big] \mathrm{d}t +\sigma \mathrm{d}W_k
\end{equation}
where the quenched disorder in this case corresponds to the presence of a non-trivial distribution of intrinsic frequencies across the oscillators, so that $h_k \sim \mu$.
The interaction kernel $K(x,y) = \sin(x-y)=\sin(x)\cos(y) - \cos(x)\sin(y)$ is separable and leads, in the thermodynamic limit, to the following NFPE of nonlinearity dimension $p=2$
\begin{equation}
\label{eq: NLFPE Kuramoto}
\partial_t \rho = - \partial_x \big( h - \mathcal{K}(x,\langle \langle \mathbf{A} \rangle \rangle) \big) \rho + \frac{\sigma^2}{2} \partial_{xx}\rho = \mathcal{L}_{\langle \langle \mathbf{A} \rangle \rangle} \rho
\end{equation}
where $\mathcal{K}(x,\langle \langle \mathbf{A} \rangle \rangle) = \sin(x) \langle \langle A_1(x) \rangle \rangle - \cos(x) \langle \langle A_2(x)\rangle \rangle $ and $\mathbf{A}(x) = \left( \cos(x),\sin(x)\right)$ represents the set of reaction coordinates for the system. 
The uniform distribution $\rho_0 =\frac{1}{2\pi}$ is always a solution of equation  \eqref{eq: NLFPE Kuramoto} corresponding to a incoherent state characterised by a set of reaction coordinates $\langle \langle \mathbf{A} \rangle \rangle_0 = (0,0)$.  We here focus on the task of assessing the response properties of the incoherent state $\rho_0$. From  \eqref{eq: matrix J}, we evaluate $\mathbf{J} = \left( \sin(x),-\cos(x)\right)$ yielding\footnote{We refer the reader to the Appendix for the details regarding the calculations regarding this section.} the microscopic Green's Function $Y_{ij}(t)$:
\begin{equation}
    Y_{ij}(t) = -\Theta(t) \int \mathrm{d}
    h \mu(h)\frac{D}{h^2+D^2} \frac{\mathrm{d}}{\mathrm{d}t}C_{ij}(t;h),
\end{equation}
where $D= \frac{\sigma^2}{2}$ and $C_{ij}(t;h)$ is the correlation function in the stationary state between observable $A_i(x)$ and $\psi_j(x;h) = A_j(x) - \frac{h}{D}\partial_x A_j(x)$. If we consider an even distribution of intrinsic frequencies $\mu(-h)=\mu(h)$ the matrix \eqref{eq: renormalisation matrix} is diagonal $P_{ij}(\omega) = P(\omega)\delta_{ij}$ and
\begin{equation}
\label{eq: function P)}
    P(\omega) = 1 - \frac{\theta D}{2}\int \frac{\mu(h)}{D^2+h^2}\left( 1 + 2i \omega C\left(\omega;h\right)\right) \mathrm{d}h
\end{equation}
where $C(\omega;h) = C_{R}(\omega;h) + i C_{I}(\omega;h)$ is the Fourier Transform of the (diagonal part of the) correlation function $C_{ij}(t;h)$ and the explicit expressions for $C_R$ and $C_I$ are in the Appendix. Singularities in the response of the reaction coordinate $A_i(x)$, due to a phase transition, are characterised by a value $\omega^* \in \mathbb{R}$ such that the matrix $P_{ij}(\omega^*)$ is not invertible, or, equivalently, such that $P(\omega^*) = 0$. Separating real and imaginary part of $P(\omega^*) = 0$ results respectively in
\begin{subequations}
    \begin{align}
    \label{eq: first condition}
        &\frac{D\theta}{2} \int \frac{\mu(h)}{h^2+D^2}\left( 1 - 2\omega^* C_I(\omega^*;h) \right) \mathrm{d}h = 1,
        \\
        \label{eq: second condition}
        & \omega^*\int \frac{\mu(h)}{h^2+D^2} C_R(\omega^*;h) \mathrm{d}h = 0.
    \end{align}
\end{subequations} 
Equation \eqref{eq: second condition} determines the critical value $\omega^* \in \mathbb{R}$ whereas \eqref{eq: first condition} identifies the corresponding transition point in the parameter space $(\theta,D)$.
An immediate solution of \eqref{eq: second condition} is $\omega^*=\omega_0 = 0$, which takes place, according to \eqref{eq: first condition}, at the critical coupling strength $ \theta^{stat}_c = 2 [ \int \frac{D}{h^2+D^2}\mu(h) \mathrm{d}h ]^{-1}$ in agreement with \cite{StrogatzMirollo1991}.
\begin{figure}[t]
    \centering
\includegraphics[width = \textwidth]{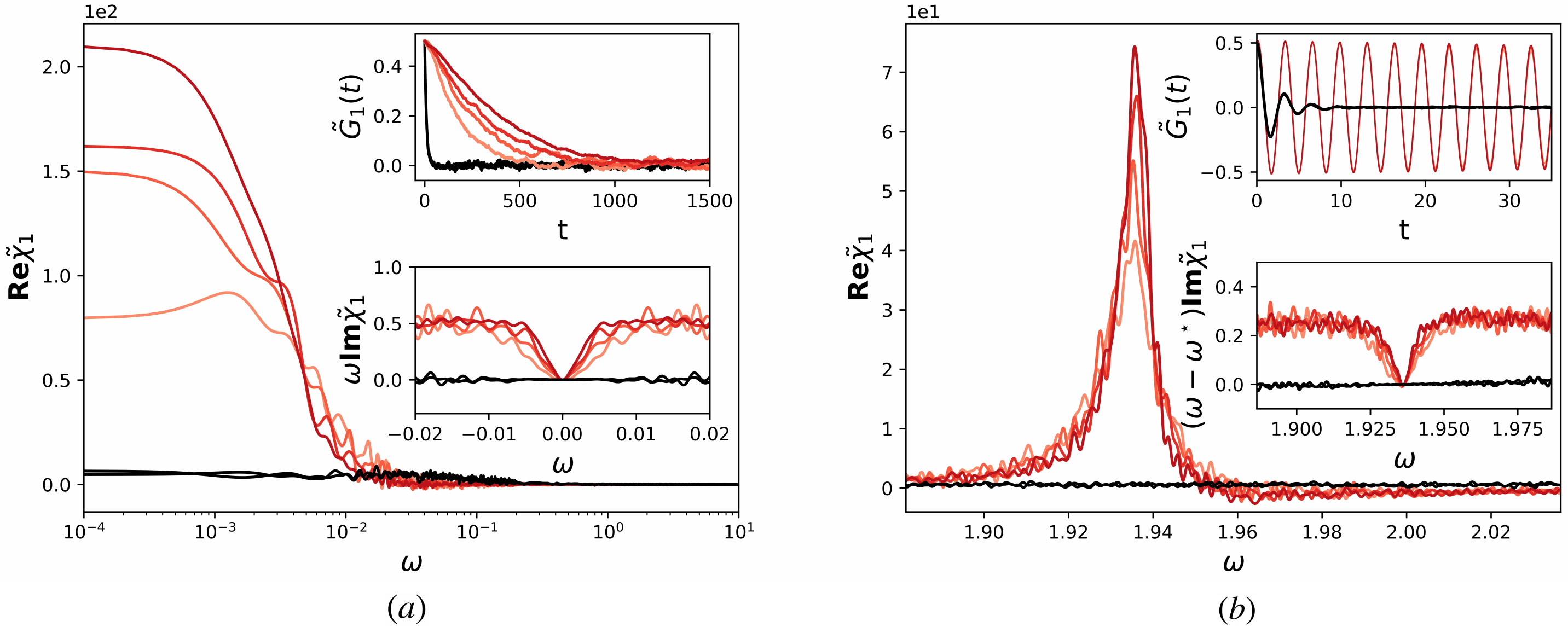}
    \caption{Real part of the susceptibility $\tilde{\chi}_1(\omega)$ associated to the reaction coordinate  $ A_1(x) $. Panel (a) refers to the homogeneous case $\mu(h) = \delta(0)$ and panel (b) to a  bimodal distribution $\mu_{h_0}(h) = \frac{1}{2} (\delta(h_0) + \delta(-h_0) )$. As the thermodynamic limit is approached, $\mathbf{Re}\tilde{\chi}_1(\omega)$ develops a diverging behaviour for $\omega = \omega_0 = 0$ (panel (a)) and $\omega = \omega^\star = \sqrt{h_0^2 - D^2}$ (panel (b)). The investigation of $\mathbf{Im} \tilde{\chi}_1$ (bottom insets) provides information on the residue associated to the pole $\omega_0=0$ of the susceptibility for the homogeneous case. The top insets show the macroscopic Green function $\tilde{G}_1(t)$.
    % As explained in the main text, the response function $\tilde{G}_1(t)$ for the bimodal case $\mu_{h_0}$ undergoes an initial linear behaviour (not shown in the figure) with positive slope proportional to the real part of the residue.
    Red (black) lines refer to settings at (away from) the phase transition. The numerical response experiments have been performed with $N = 6,12,24,48 \times 10^3$, see more details in the Appendix.}
    \label{fig:my_label}
\end{figure}
This setting corresponds to a static phase transition, characterised by a vanishing pole $\omega_0=0$ of the susceptibility and a lack of oscillations in the system, see panel (a) of Figure \ref{fig:my_label} where we have investigated this scenario for $\mu_0(h)= \delta(h)$. The introduction of a distribution of natural frequencies $\mu$ brings the system to a nonequilibrium regime and it is interesting to investigate whether a dynamical phase transition, characterised by a pair of poles $\omega_1 = - \omega_2 = \omega^* > 0$, can arise in the system. This involves the study of the dynamical properties encoded in $C(\omega;h)$. 
Further insight can be gained by observing that, if \eqref{eq: second condition} is satisfied for $\omega^* >0$, the following equation holds, see details in the Appendix,
\begin{equation}
\label{eq: real part Kuramoto}
     \int \frac{\omega^* \pm h}{D^2 + (\omega^* \pm h)^2} \mu(h)\mathrm{d}h = 0,
\end{equation}
which is a known result for the Kuramoto model \cite{GuptaCampaRuffo2014}. In particular, no dynamical phase transition can take place if $\mu(h)$ is unimodal as no $\omega^* \neq 0$ satisfying \eqref{eq: real part Kuramoto} exists. 
We here consider, as in \cite{Bonilla1992}, a bimodal distribution $\mu_{h_0}(h) = \frac{1}{2}(\delta(h-h_0)+\delta(h+h_0))$. In this setting, equation \eqref{eq: second condition} admits, further to the static transition scenario, a new solution $\omega^*= \sqrt{h_0^2 - D^2}$ when $h_0>D$. According to \eqref{eq: first condition}, this dynamical phase transition scenario takes place at the transition point $\theta_{dyn} = 4D$. Panel (b) of Figure \ref{fig:my_label} confirms that, at the phase transition, response properties break down as the susceptibility develops a singular resonant behaviour for  $\omega = \omega^*$ as $N \to \infty$, resulting in greatly amplified and long lasting oscillations of the macroscopic Green's function $\tilde{G}_i(t)$. 
\subsection{Response properties of the finite system}
Following \cite{Marconi2008}, we evaluate response properties of the finite system by performing $n$ simulations of equations \eqref{eq : Kuramoto model} where the initial conditions are sampled according to $\rho_0 = \frac{1}{2 \pi}$ and where at time $t = 0$ we apply a macroscopic perturbation $F(x;h)=h \to F(x;h)=h + \varepsilon X(x)\delta(t)$. The macroscopic Green's function $\tilde{G}_1(t)$ is obtained as the average over the $n$ ensemble members of $\langle \langle A_1 \rangle \rangle_1(t)$  which we estimate as $\frac{1}{N}\sum_{i=1}^N A_1(x_i(t))$. We remark that, since the unperturbed measure is uniform, a constant perturbation $X(x) = 1$ would result in a vanishing response as $G_1(t) \equiv 0$, see \eqref{eq: microscopic Green Function G}. We have here chosen a  non trivial state dependent perturbation $X(x) \propto \sin(x)$. We note that such term has been used to model excitability features in Kuramoto-like models for brain dynamics \cite{Buendia2022}.
The numerical experiments indicate that the finite system susceptibility can be written, at the phase transition, as 
\begin{equation}
\label{eq: mollified susceptibility}
     \tilde{\chi}_1(\omega) = 
     \begin{cases} 
    \frac{\kappa}{\omega + i \gamma(N) } + \psi(\omega) & \text{if $\omega^*=0$}
    \\
    \frac{\kappa}{\omega- \omega^* + i \gamma(N)}  - \frac{\kappa^*}{\omega+ \omega^* + i \gamma(N)} + \psi(\omega) & \text{if $\omega^* >0$}
    \end{cases}
\end{equation} where $\psi(\omega)$ is an analytic function in the upper complex $
\omega$ plane and the residue $\kappa$ characterises the strength of the divergence. The residue is related to the properties of $P(\omega)$ and can be found as $\kappa = \left( \theta P'(\omega^*) \right)^{-1}$, see Appendix for more details.
\begin{figure}
    \centering
    \includegraphics[width=\textwidth]{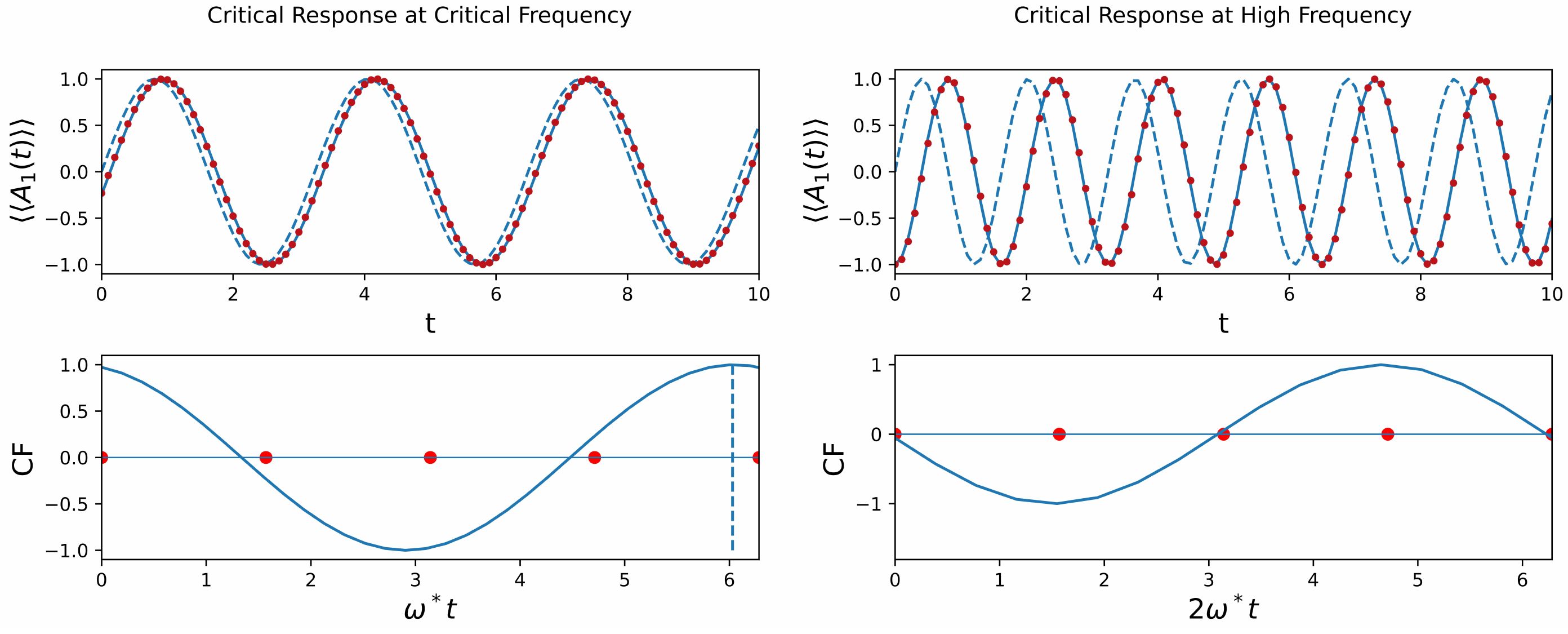}
    \caption{Top panels: Response of the reaction coordinate $A_1$ to a sinusoidal forcing $T(t)=\sin(\omega t)$, with $\omega = \omega^*$ (left) and $\omega = 2\omega^*$ (right) . The dashed line corresponds to $T(t)$, the continuous one to the analytical result and the red dots to the numerical response evaluated as $\tilde{G}_1 \star T$ with $\tilde{G}_1(t)$ obtained from the numerical simulations for $N=48 \times 10^3$. The amplitude of the numerical response has been re-scaled by its maximum value. Bottom panel: Correlation function (CF) between $T(t)$ and the numerical response as a function of the angle $\omega t$. Its maximum value yields an estimate of the phase shift $\phi$. The vertical dashed line corresponds to the analytical value of $\phi$. Red dots correspond to multiples of $\frac{\pi}{2}$ and are just a visual aid.}
    \label{fig:phase shift critical}
\end{figure}
We remark that the finiteness of the system results in a mollifying effect of the singularity into a resonant behaviour, where the poles are slightly shifted as $\pm \omega^* \to \pm \omega^* + i \gamma(N)$, with $\gamma(N) \to 0$ as $N\to +\infty$. In the homogeneous case the residue turns out to be completely imaginary and equal to $\kappa = \frac{i}{2}$, whereas for the bimodal case it results in $\kappa = \kappa_R + i \kappa_I = - \frac{D}{4 \omega^*} + \frac{i}{4}$. In the former case, one expects that, as $N \to +\infty$, the function $\omega \mathbf{Im}\tilde{\chi}_1(\omega)$ will be constant and equal to $|\kappa| = \frac{1}{2}$ in the proximity of the pole and to quickly drop to zero for $\omega =0$. A visual inspection of the bottom inset of panel (a) of Figure \ref{fig:my_label} provides a clear evidence of such theoretical prediction.
\\
A fundamentally different situation occurs in the latter case as the residue has a non-vanishing real part leading to a distortion of the resonance at $\omega^*$ as $\mathbf{Im}\tilde{\chi}(\omega^*) \neq 0$ and, thus, affecting the way the amplitude of the oscillations of the response track the forcing signal. This is more easily understood when considering a sinusoidal driving forcing $T(t) = \sin(\hat{\omega} t)$. The response (in the time domain) can be obtained by performing an inverse Fourier transform of \eqref{eq: response in frequency with macroscopic susceptibility} and results in
\begin{equation}
     \langle \langle A_1 \rangle \rangle_1(t;\hat \omega) = |\tilde{\chi}(\hat \omega)|\sin\left( \hat \omega t+ \arctan\left( \frac{\mathbf{Re}\tilde{\chi}(\hat \omega)}{\mathbf{Im}\tilde{\chi}( \hat \omega)}\right) - \frac{\pi}{2} \right) := |\tilde{\chi}(\hat \omega)|\sin\left( \hat \omega t+ \phi \right)
\end{equation}
The amplitude of the response is given by the amplitude of the macroscopic susceptibility. We expect that driving frequencies around the pole $\omega^*$ would destabilise the system resulting in greatly amplified oscillations, whereas the system would be much more resilient for higher driving frequencies. Moreover, if the imaginary part of the susceptibility were to vanish,  the response would feature either no phase shift as $\phi = 0$ or phase reversal as $\phi = - \pi$. At the critical frequency $\hat{\omega} = \omega^*$ we have, from \eqref{eq: mollified susceptibility}, that the response is given by  $\langle \langle A_1 \rangle \rangle_1(t;\omega^*) = \frac{|\kappa|}{\gamma(N)}\sin\left(\omega^* t + \phi \right)$ with $\phi = \arctan(-\frac{\kappa_I}{\kappa_R}) - \frac{\pi}{2}= \arctan(\frac{\omega^*}{D})-\frac{\pi}{2}$. The non vanishing real part of the residue yields a spontaneous phase shift of the response to perturbations as confirmed by Figure \ref{fig:phase shift critical} (left panels). We remark that this provides a quantitative estimate, at the phase transition, of the phase shift phenomenon observed (in deterministic settings) in \cite{Petkoski2012}. On the other hand, if the system is driven at higher frequency, e.g. $\hat{\omega}= 2\omega^*$, we expect the real part of the susceptibility to decay faster than the imaginary part \cite{Lucarini2005} and the response to be in quadrature with the forcing, i.e. $\langle \langle A_1 \rangle \rangle_1(t;2\omega^*) \propto \cos(2\omega^* t)$, as confirmed by Figure \ref{fig:phase shift critical}  (right panels).
\subsection{\textbf{Conclusions}}
\label{sec: conclusion}
In this paper we have considered the relevant issue of the identification of a set of reaction coordinates $\{ A_i \}$ for the thermodynamic limit of ensembles of interacting agents in the general case of pairwise interaction and quenched disorder in the dynamics. Given a general class of interaction structure among the agents, we have provided a framework to define the $\{ A_i \}s$ and construct a nonequilibrium thermodynamical formulation of the macroscopic description of the system. These reaction coordinates provide a parametrization of the stationary measure and fully characterise the linear response of the system to perturbations. Such a model reduction strategy is particularly useful as it allows to study not only smooth regimes of the response but also critical behaviours, characterised as an emergence of singular poles of the complex valued susceptibility of the system. The analysis in terms of the reaction coordinates of the paradigmatic Kuramoto model allows us to show (a) that our framework encompasses  known results on the stability properties of the incoherent state and (b) that the time dependent properties of the onset of synchronisation can be quantitatively investigated through the susceptibility function describing the response of the system to general driving forces.
\\
Our results are exact in the case of separable interactions. However, non separable kernels are also found in applications. If $\mathbf{K}(\mathbf{x},\mathbf{y})$ is a smooth function, an expansion as in \eqref{eq: sep kernel2} still holds by using a Schauder decomposition \cite{LindenstraussTzafriri1977} but it contains in general an infinite number of terms, resulting in an infinite number of observables $\{  A_i \} $ with a clear loss of  dimension reduction of the system. However, there exist numerous rigorous techniques to approximate in a controlled manner a general kernel $\mathbf{K}(\mathbf{x},\mathbf{y})$ in terms of suitable separable kernels, see \cite[Ch. 11]{Kress2014lie}\cite{DELLWO1995}. Such approximating methods would be the basis for the development of phenomenological theories of interacting systems with suitable separable interactions for which the identification of a finite number of observables can be performed exactly. 
\\
On the one hand, the investigation of mean field models is very relevant as it provides a way of gaining a mathematical and analytical understanding of the smooth and critical features of  ensembles of interacting agents. On the other hand, the lack of an underlying geometrical structure might pose a too restrictive limitation to the applications of these models. Future research in this direction will involve the generalisation of this framework to systems with a microscopic network structure \cite{QuininaoTouboul2015} and in particular to all systems amenable to a description in terms of graphon \cite{LovaszGraphonBook, Coppini2022} and graphop theory \cite{Kuehn_2020,Gkogkas2022}. We envision that this framework can be extended to include the physically relevant cases where the dynamics is not assumed to be overdamped, in the direction of understanding nonequilibrium ensembles in statistical mechanics \cite{G14} and  to non-Markovian interacting particles~\cite{DuongPavliotis2018}, where appropriate Markovian closures have to be considered \cite{pavliotisbook2014,santos2021}.
\ack
VL acknowledges the support received by the European Union’s Horizon 2020 research and innovation program through the project TiPES (Grant Agreement No. 820970) and Marie Curie ITN CriticalEarth (Grant Agreement No. 956170). The work of GP was partially funded by the EPSRC, grant number EP/P031587/1, by J.P. Morgan Chase \& Co through a Faculty Research Award 2019 and 2021 and by Advanced ERC grant Machine-Aided General Framework for Fluctuating Dynamic Density Functional Theory. NZ has been supported by the Wallenberg Initiative on Networks and Quantum Information (WINQ). 

\appendix

\section{Dynamical Properties of the Kuramoto model}
We provide here the detailed calculations leading to the results reported in the main text regarding the Kuramoto model. This model describes an ensemble of one dimensional  ($M=1$) phase oscillators on the torus $\mathbb{T}^1 = [0,2\pi]$ whose dynamics is determined by the following Stochastic Differential Equations
\begin{equation}
\label{eq : Kuramoto model appendix}
    \mathrm{d}x_k = \big[ h_k -\frac{\theta}{N} \sum_{j=1}^N \sin\left( x_k - x_j \right) \big] \mathrm{d}t +\sigma \mathrm{d}W_k, \quad k = 1, \dots, N
\end{equation}
where $h_k \sim \mu$, with $\mu(-h)= \mu(h)$, represents the intrinsic frequencies of each oscillator and the symbols are defined as in the main text. The interaction kernel $K(x,y) = \sin(x-y)=\sin(x)\cos(y) - \cos(x)\sin(y)$ is separable and leads, in the thermodynamic limit $N \to + \infty$ \cite{DaiPra,PraHollander}, to the following nonlinear nonlocal Fokker Planck equation of nonlinearity dimension $p=2$
\begin{equation}
    \label{eq appendix: NLFPE Kuramoto}
\partial_t \rho = - \partial_x \big( h - \mathcal{K}(x,\langle \langle \mathbf{A} \rangle \rangle) \big) \rho + \frac{\sigma^2}{2} \partial_{xx}\rho = \mathcal{L}_{\langle \langle \mathbf{A} \rangle \rangle} \rho
\end{equation}
where $\mathcal{K}(x,\langle \langle \mathbf{A} \rangle \rangle) = \sin(x) \langle \langle A_1(x) \rangle \rangle - \cos(x) \langle \langle A_2(x)\rangle \rangle $ where $\mathbf{A}(x) = \left( \cos(x),\sin(x)\right)$ represents the set of reaction coordinates for the Kuramoto model. It is easy to verify that the uniform solution $\rho_0 =\frac{1}{2\pi}$ is always a stationary solution of equation  \eqref{eq appendix: NLFPE Kuramoto} corresponding to a disordered state characterised by a set of observables $\langle \langle \mathbf{A} \rangle \rangle_0 = (0,0)$. Moreover, this stationary solution $\rho_0$ satisfies $\mathcal{L}_{0} \rho_0 = 0$ where 
\begin{equation}
\label{eq: parametrised operator}
\mathcal{L}_{0} := \mathcal{L}_{\langle \langle \mathbf{A}\rangle \rangle_0}   = -h \partial_x + \frac{\sigma^2}{2}\partial_{xx}
\end{equation}
Below we investigate the response properties of the disordered state and detect phase transitions of the system by looking at the singularities of the susceptibility of the system originating from non invertibility properties of the matrix $P_{ij}(\omega)$, see equation $(12)$ in the main text. 
We evaluate the matrix $\mathbf{J} \in \mathbb{R}^2$ (here a vector since $M=1$)  
\begin{equation}
\mathbf{J} = \frac{\partial \mathcal{K}}{\partial \langle \langle \mathbf{A} \rangle \rangle} = \left(\sin(x), - \cos(x)\right)
\end{equation}Considering that $\partial \mathbf{J}(x) = \mathbf{A}(x)$, the microscopic Green's function $Y_{ij}(t)$ can be written as $Y_{ij}(t) = \int \mathrm{d}h  \mu(h) Y_{ij}(t;h)$ where
\begin{equation}
Y_{ij}(t;h) = \Theta(t)\int_0^{2\pi} \frac{\mathrm{d}x }{2\pi}A_i(x)e^{t \mathcal{L}_{0}} A_j(x)
\end{equation}Now, it is easy to verify that 
\begin{equation}
\mathbf{A} = - \frac{D}{h^2+D^2}\mathcal{L}_{0} \boldsymbol{\psi}
\end{equation}where $D=\frac{\sigma^2}{2}$ and 
\begin{equation}
\boldsymbol{\psi}(x;h) = 
\begin{pmatrix}
\cos(x) + \frac{h}{D}\sin(x) \\
\sin(x) - \frac{h}{D} \cos(x)
\end{pmatrix} = \mathbf{A}(x) - \frac{h}{D}\partial_x \mathbf{A}(x)
\end{equation}
The microscopic Green Function can then be written as 
\begin{equation}
\label{eq: intrinsic susceptibility}
\begin{split}
Y_{ij}(t;h) &= - \frac{D}{h^2+D^2}\Theta(t) \int_{0}^{2\pi} \frac{\mathrm{d}x}{2\pi} A_i(x) e^{t \mathcal{L}_{0}} \mathcal{L}_{0} \psi_j(x;h) =\\ 
& = -\frac{D}{h^2+D^2} \Theta(t) \frac{\mathrm{d}}{\mathrm{d}t} \int_{0}^{2\pi} \frac{\mathrm{d}x}{2\pi} A_i(x) e^{t \mathcal{L}_{0}} \psi_j(x;h) = \\
&:= - \frac{D}{h^2+D^2}\Theta(t) \frac{\mathrm{d}}{\mathrm{d}t}C_{ij}(t;h)
\end{split}
\end{equation}where $C_{ij}(t;h)$ is the correlation function between observable $A_i(x)$ and $\psi_j(x;h)$. The corresponding microscopic susceptibility is then given by 
\begin{equation}
Y_{ij}(\omega;h) = \int_{-\infty}^{+\infty} e^{i\omega t}Y_{ij}(t)\mathrm{d}t=\frac{D}{h^2+D^2} \left( C_{ij}(t=0;h)+i\omega C_{ij}(\omega;h)\right)
\end{equation}
where $C_{ij}(\omega;h) = \int_{0}^{+\infty} e^{i\omega t}C_{ij}(t;h)\mathrm{d}t $ is the (one sided) Fourier Transform of $C_{ij}(t;h)$. Since the correlation function at time $t=0$ is $C_{ij}(t=0;h) = \frac{1}{2}\delta_{ij} - \frac{h}{2D}(1-\delta_{ij})$ the microscopic susceptibility associated to $Y_{ij}(t)$ is
\begin{equation}
\label{eq: intrinsic susc Kuramoto}
 Y_{ij}(\omega) = \int Y_{ij}(\omega;h)\mu(h)\mathrm{d}h = \frac{D}{2}\delta_{ij} \int \frac{\mu(h)}{D^2+h^2}\mathrm{d}h + i \omega D \int  \frac{C_{ij}(\omega;h)}{h^2+D^2} \mu(h)\mathrm{d}h
\end{equation}where we have used the fact that the distribution of frequencies is even. The matrix $P_{ij}(\omega)$ describing the response to the observable $\langle \langle A_i(x) \rangle \rangle$ is given by 
\begin{equation}
\label{eq: renormalisation Kuramoto}
P_{ij}(\omega) = \delta_{ij} - \theta Y_{ij}(\omega) = \left( 1 - \frac{\theta D}{2} \int \frac{\mu(h)}{D^2+h^2}\mathrm{d}h \right)\delta_{ij} - i \theta \omega D \int \frac{{C}_{ij}(\omega;h)}{h^2+D^2} \mu(h)\mathrm{d}h
\end{equation}
\subsubsection{Evaluation of the Correlation Function}
The correlation function $C_{ij}(t;h)$ defined in \eqref{eq: intrinsic susceptibility} is completely determined by the spectral properties of the operator $\mathcal{L}_{0}$ which is a linear differential operator with constant coefficients, see equation \eqref{eq: parametrised operator}. As such, its spectral features on the space of functions that are square integrable with respect to the invariant measure $L^2(\mathbb{T};\rho_0)$ is easily found to be given by the eigenfunctions $\phi_k = e^{ikx}$ with relative eigenvalue $\lambda_k = - ihk - k^2 D$ where $k=0, \pm 1, \pm 2, \dots$ . We remark that the eigenfunctions $\phi_k$ are orthonormal in $L^2(\mathbb{T};\rho_0)$ since $\langle \phi_k | \phi_{k'} \rangle_0 = \int_0^{2\pi} \phi_k^*(x) \phi_{k'}(x) \rho_0 = \frac{1}{2\pi}\int_0^{2\pi} e^{i (k' - k)x} = \delta_{kk'}$ where $\langle \cdot | \cdot \rangle_0$ represents the $L^2(\mathbb{T},\rho)$ scalar product. 
We can then obtain a spectral decomposition \cite{EngelNagel2000,ChekrounJSPI} of the operator $e^{t \mathcal{ L}_0}$ as
\begin{equation}
    e^{t\mathcal{L}_0} = \sum_{k = - \infty}^{+\infty} e^{t \lambda_k} \Pi_k
\end{equation}
where $   \Pi_k = | \phi_k \rangle \langle \phi_k | $  is the projector onto the eigenspace in $L^2(\mathbb{T};\rho_0)$ relative to the eigenvalue $\lambda_k$. We then obtain 
\begin{equation}
\begin{split}
C_{ij}(t;h) &= \sum_{k = - \infty}^{+ \infty} e^{t \lambda_k} \int \mathrm{d}x A_i(x) \Pi_k \psi_j(x;h) =
 \sum_{k = - \infty}^{+ \infty} e^{t \lambda_k} \langle \frac{A_i}{\rho_0} | \phi_k \rangle_0 \langle \phi_k | \psi_j \rho_0 \rangle_0 := \sum_{k = - \infty}^{+ \infty} e^{t \lambda_k}  \alpha_{ij}^{(k)}(h)
\end{split}
\end{equation}
Since $\lambda_{-k} = \lambda_k^*$ and $\phi_{-k}= \phi_k^*$,  the coefficients $\alpha_{ij}^{(k)}$ satisfy $\alpha_{ij}^{(-k)} = \left(\alpha_{ij}^{(k)}\right)^*$ and the correlation function can be written as 
\begin{equation}
C_{ij}(t;h) = 2 \mathbf{Re} \sum_{k=1}^{+\infty} e^{\lambda_k t}\alpha_{ij}^{(k)} 
\end{equation}where the mode $k=0$ does not contribute to the correlation function, as $\Pi_0$ projects onto the invariant measure $\rho_0 = \frac{1}{2\pi}$ yielding $a_{ij}^{(0)} = 0$. Given the particular structure of the coupling kernel $K(x,y) = sin(x-y)$ selecting the reaction coordinates $\mathbf{A}(x)$, it is possible to show that only the first $k=1$ contributes to the correlation functions, since $\alpha_{ij}^{(k)} = 0$ for $k=2,3,\dots$ . Indeed a simple but lengthy calculation shows that 
\begin{equation}
\label{eq: coefficients}
\alpha_{ij}^{(k)} = \frac{\delta_{k1}}{4} \left(\delta_{ij}  - \frac{h}{D}\left(1 - \delta_{ij} \right) S_{ij} + i \left(  \left(  \left(1- \delta_{ij} \right) A_{ij} - \frac{h}{D}\delta_{ij}\right) \right) \right)
\end{equation}where $\mathbf{S} = \begin{pmatrix} 0 & 1 \\ 1 & 0\end{pmatrix}$ and $\mathbf{A}= \begin{pmatrix} 0 & -1 \\ 1 & 0
\end{pmatrix}$ are two antidiagonal matrices describing the interaction between the observables $A_i$ and $\psi_j$ with $i \neq j$. 
The correlation function can then be written as 
\begin{equation}
C_{ij}(t;h) =  2 \mathbf{Re} \left( e^{\lambda_1 t} \alpha_{ij}^{(1)} \right) = \frac{\delta_{ij}}{2} e^{-Dt} \left( \cos(ht) - \frac{h}{D}\sin(ht) \right) + c_{ij}(t;h)
\end{equation}where $c_{ij}(t;h)$ is an off diagonal contribution and is an odd function of the intrinsic frequency, $c_{ij}(t;-h)=-c_{ij}(t;h)$. Since we have assumed that the intrinsic frequency distribution is even, we observe that equation \eqref{eq: renormalisation Kuramoto} shows that $c_{ij}(t;h)$ provides a vanishing contribution to the matrix $P_{ij}(\omega)$ and in the following calculations it will be neglected. 
By taking the (one sided) Fourier transform we obtain
\begin{equation}
\label{eq: FT correlation transform Kuramoto}
C_{ij}(\omega;h) = \frac{\delta_{ij}}{2} \frac{-D+h^2/D + i \omega}{\omega^2 - (D^2+h^2)+2i \omega D} :=  C(\omega;h)\delta_{ij},
\end{equation}where we have defined
\begin{subequations}
\begin{align}
\label{eq: def of C}
C(\omega;h) &=  C_R(\omega;h) + i C_I(\omega;h) 
\\
\label{eq: real part C}
C_R(\omega;h) &= \frac{D^2+h^2}{2D} \frac{\omega^2+D^2-h^2}{\left(\omega^2 - \left( D^2 + h^2 \right)  \right)^2 + 4 \omega^2 D^2}, \\
\label{eq: imaginary part C}
C_I (\omega;h)  &= \frac{ \omega}{2}  \frac{\omega^2 +D^2- 3h^2}{\left(\omega^2 - \left( D^2 + h^2 \right)  \right)^2 + 4 \omega^2 D^2}.
\end{align}
\end{subequations}
From equation \eqref{eq: renormalisation Kuramoto} and \eqref{eq: FT correlation transform Kuramoto} we note that the renormalisation matrix is diagonal $P_{ij}(\omega) = P(\omega)\delta_{ij}$, with
\begin{equation}
\label{eq: equation for P}
     P(\omega) = 1 - \frac{\theta D}{2}\int \frac{\mu(h)}{D^2+h^2}\left( 1 + 2i \omega C\left(\omega;h\right)\right) \mathrm{d}h.
\end{equation}
\subsubsection{Phase transitions for the Kuramoto model}
As explained in the main text, phase transitions are associated to settings where the matrix $P_{ij}(\omega^*)$ is not invertible for $\omega^* \in \mathbb{R}$. Given equation \eqref{eq: equation for P}, this is equivalent to the condition $P(\omega^*) = 0$. Now we separate real and imaginary part of the above equation and, considering  \eqref{eq: def of C} and \eqref{eq: equation for P} we get, respectively,
\begin{subequations}
    \begin{align}
        \label{eq: first condition appendix}
\frac{D\theta}{2}& \int \frac{\mu(h)}{h^2+D^2}\left( 1 - 2\omega^* C_I(\omega^*;h) \right) \mathrm{d}h = 1,\\
\label{eq: second condition appendix}
\omega^* &\int \frac{\mu(h)}{h^2+D^2} C_R(\omega^*;h) \mathrm{d}h = 0.
    \end{align}
\end{subequations}
Equation \eqref{eq: second condition appendix} provides the value of $\omega^*$ in terms of the parameters of the system and the properties of the quenched disorder distribution $\mu$ whereas \eqref{eq: first condition appendix} gives the associated transition point, that is a relationship among the parameters of the system $(\theta,D)$ and the properties of $\mu$. In general, we can identify two typical scenarios of phase transitions: static phase transitions associated to a single simple pole $\omega^*=0$ and dynamic phase transitions associated to an opposite pair of pole $\omega_1 = - \omega_2 = \omega^* > 0$. 
\subsubsection{Static phase transitions}
It is immediate to see that $\omega^* = \omega_0 = 0$ is a solution of \eqref{eq: second condition appendix}. Evaluating \eqref{eq: first condition appendix} for $\omega^*=\omega_0 = 0$ yields
\begin{equation}
1 - \frac{\theta D}{2}\int \frac{\mu(h)}{h^2+D^2} \mathrm{d}h = 0
\end{equation}
from which we obtain the critical value of the coupling strength
\begin{equation}
\label{eq: critical strength}
\theta^{stat}_c = 2 \bigg[ \int \frac{D}{h^2 + D^2}\mu(h)\mathrm{d}h \bigg]^{-1} 
\end{equation}We remark that this phase transition is characterised by a vanishing simple pole $\omega_0 =0$, resulting in a static transition where the static susceptibility $\tilde{ \boldsymbol{\chi}}(0) = \left( \mathbf{P}^{-1}\boldsymbol{\chi} \right)\left( 0\right) $ diverges.
\subsubsection{Dynamic phase transitions: properties of $\mu$}
The investigation of the onset of a dynamical phase transition is more complicated as it requires the study of frequency dependent properties of $C(\omega;h)$. Since we are considering a dynamic phase transition characterised by $\omega^* \neq 0$, equation \eqref{eq: second condition appendix} can be written as
\begin{equation}
\label{eq: real part dyn}
    \int \frac{\mu(h)}{h^2+D^2} C_R(\omega^*;h) \mathrm{d}h = \frac{1}{2D}\int \frac{\omega^2+D^2-h^2}{\left(\omega^2 - \left( D^2 + h^2 \right)  \right)^2 + 4 \omega^2 D^2} \mu(h) = 0.
\end{equation}
where in the last equality we have used \eqref{eq: real part C}. 
We can prove that \eqref{eq: real part dyn} implies a known result for the Kuramoto model by noticing that such equation contains the notion that, if $\omega^* \neq 0$ is a solution, then $-\omega^*$ is also a solution. Indeed it is easy to show that for $\omega \neq 0$ the following identity holds
\begin{equation}
 \frac{\omega^2+D^2-h^2}{\left(\omega^2 - \left( D^2 + h^2 \right)  \right)^2 + 4 \omega^2 D^2} = \frac{1}{2\omega} \left(  \frac{\omega+h}{D^2 + (\omega+h)^2} + \frac{\omega - h}{D^2 + (\omega -h )^2}\right)
\end{equation}
so that \eqref{eq: real part dyn} implies that
\begin{equation}
\int  \left( \frac{\omega+h}{D^2 + (\omega+h)^2} + \frac{\omega - h}{D^2 + (\omega -h )^2} \right)  \mu(h)\mathrm{d}h = 0.
\end{equation}
Considering that $\mu(h)$ is an even function the previous equation implies that
\begin{equation}
\label{eq: Kuramoto condition}
    \int \frac{\omega^* \pm h}{D^2 + (\omega^*\pm h)^2} \mu(h) \mathrm{d}h = 0,
\end{equation}
which is a known formula for the Kuramoto model \cite{GuptaCampaRuffo2014}. The above equation shows that the existence of a nonvanishing singularity $\omega^*$ is closely related to the properties of the intrinsic frequency distribution $\mu(h)$. In particular, if $\mu(h)$ is a unimodal function then there does not exist a $\omega^* \neq 0$ satisfying \eqref{eq: Kuramoto condition}. In other words, an even unimodal distribution cannot support the development of a dynamic phase transition.
\subsubsection{Dynamic phase transition for a bimodal frequency distribution}
The simplest setting that can support a dynamic phase transition scenario is represented by a bimodal distribution
\begin{equation}
\mu_{h_0}(h) = \frac{\delta(h+h_0)+\delta(h-h_0)}{2}.
\end{equation}
In this case, \eqref{eq: real part dyn} becomes simply
\begin{equation}
\label{eq: condition bimodal}
    C_R(\omega^*;h_0) = 0
\end{equation}
that, given \eqref{eq: real part C}, yields a solution $\omega^* = \sqrt{h_0^2 - D^2} $. This solution $\omega^*$ is real only if $h_0 > D$, that is, a sufficiently large separation of the peaks of the bimodal distribution is needed, as observed in \cite{Bonilla1992,AcebronBonilla2005}. 
The associated phase transition point is given by \eqref{eq: first condition appendix}, which reads for a bimodal distribution
\begin{equation}
\label{eq: bimodal phase trans cond}
    \frac{\theta D}{2} \left(1 - 2\omega^* C_I(\omega^*;h_0)  \right) = h_0^2 +D^2
\end{equation}
We first evaluate the denominator of $C_I(\omega^*;h_0)$
\begin{equation}
    \left( \omega_*^2 - \left( D^2 + h_0^2 \right)  \right)^2 + 4 \omega_*^2 D^2 =  (2D^2)^2 + 4 D^2 \omega_*^2 = 4D^2(D^2+\omega_*^2)
\end{equation}
 where for the sake of notation clarity we have defined $\omega_*^2 := (\omega^*)^2 = h_0^2 -D^2$. We can then write from \eqref{eq: imaginary part C} 
 \begin{equation}
 \label{eq: condition bimodal on imaginary part C}
     C_I(\omega^*;h_0) = \frac{\omega^*}{2} \frac{\omega_*^2 + D^2 -3h_0^2}{ \left( \omega_*^2 - \left( D^2 + h_0^2 \right)  \right)^2 + 4 \omega_*^2 D^2} = -  \omega^* \frac{h_0^2}{4D^2(D^2+\omega_*^2)} = - \frac{\omega^*}{4D^2}
 \end{equation}
 where in the last equality we have used the fact that by definition of $\omega^*$ we have that $D^2 + \omega_*^2 = h_0^2$. Finally, equation \eqref{eq: bimodal phase trans cond} results in 
 \begin{equation}
     \begin{split}
    \frac{\theta D}{2} \left(1 + \frac{\omega^2}{2D^2} \right) &= h_0^2 + D^2, \\
     \frac{\theta D}{2} \frac{2D^2+ \omega_*^2}{2D^2} &= h_0^2 +D^2, \\
     \frac{\theta D}{2} \frac{h_0^2+D^2}{2D^2} &= h_0^2 +D^2, \\
     \frac{\theta D}{2}\frac{1}{2D^2} &= 1,
     \end{split}
 \end{equation}
 yielding eventually the critical phase transition point $\theta_c^{dyn} = 4D$.
\subsection{Characterisation of phase tranisitions: residues}
 The response properties of the system to external perturbations are characterised by the total susceptibility 
\begin{equation}
    \tilde{\chi}_i(\omega) =  \left( \mathbf{P}^{-1} \boldsymbol{\chi} \right)_i\left(\omega\right) = P^{-1}(\omega) \chi_{i}(\omega),
\end{equation}
where in the last equality we have used the fact that $P_{ij}(\omega) = P(\omega) \delta_{ij}$, with $P(\omega)$ given by \eqref{eq: equation for P}, when the distribution of frequencies $\mu(h)$ is an even function. Since the microscopic Green's Function $G_i(t)$ is given by, see equation $(10a)$ in the main text, 
\begin{equation}
    G_i(t) = \int \mathrm{d}h \mu(h)\int \mathrm{d}x A_i(x) e^{t\mathcal{L}_{0} } \mathcal{L}_p \rho_0,
\end{equation}
where $\mathcal{L}_p \rho_0 = - \partial_x \left( X(x) \rho_0 \right)$, we see that a homogeneous perturbation $X(x) = const$ would result in a null response $G_i(t)=0$, and consequently $\chi_i(\omega)=0$, given that the stationary measure is the uniform distribution $\rho=\frac{1}{2\pi}$. We have chosen to perform the numerical response simulations with $X(x) = - \sin(x) = A_2(x)$, so that the microscopic Green's Function is  $G_i(t) = Y_{i1}(t)$ and the corresponding susceptibility is $\chi_i(\omega) = Y_{i1}(\omega)$. 
\subsubsection{Residue for static phase transitions}
A static phase transition is characterised by a simple pole at $\omega=\omega^*=0$, so that the critical response of the reaction coordinate $A_1(x)$ can be written as 
\begin{equation}
\label{eq: susc as pole}
\tilde{\chi}_1(\omega) = \frac{Y_{11}(\omega)}{P(\omega)} =  \frac{Y_{11}(0)}{P'(0)\omega} + \psi(\omega) =  \frac{1/\theta}{P'(0)\omega} + \psi(\omega) := \frac{\kappa}{\omega} + \psi(\omega),
\end{equation}
where $\psi(\omega)$ is an analytic function in the upper complex $\omega^*$ plane and we have used the fact that, at a static phase transition, $P(0) = 0$ and $Y_{11}(0)= (1-P(0))/\theta = 1/\theta$. The quantity $\kappa =  \frac{1/\theta}{P'(0)}$ represents the residue of the pole $\omega^*=0$ and determines the properties of the singularity. From \eqref{eq: equation for P} we evaluate the derivative of $P(\omega)$
\begin{equation}
\label{eq: derivative of P}
    P'(\omega) = - i \theta D \int \frac{\mu(h)}{D^2+h^2} \left( C(\omega;h) + \omega C'(\omega;h)\right) \mathrm{d}h
\end{equation}
For a static phase transition we are interested in 
\begin{equation}
    P'(0) = -  i \theta D \int \frac{\mu(h)}{D^2+h^2}C(0;h) \mathrm{d}h = - \frac{i \theta}{2} \int \mu(h) \frac{D^2-h^2}{\left( D^2+h^2 \right)^2} \mathrm{d}h
\end{equation}
where in the last equality we have used that $C(0;h)=  C_R(0;h) +i C_I(0;h) = C_R(0;h) = \frac{1}{2D}\frac{D^2-h^2}{D^2+h^2}$. 
The residue is thus 
\begin{equation}
\label{eq: static residue}
    \kappa = \frac{2i}{\theta^2 \int \mu(h) \frac{D^2-h^2}{\left( D^2+h^2 \right)^2} \mathrm{d}h} = i \frac{\left( \int \frac{D}{h^2+D^2}\mu(h)\mathrm{d}h\right)^2}{2\int  \frac{D^2-h^2}{\left( D^2+h^2 \right)^2}\mu(h) \mathrm{d}h }.
\end{equation}
where we have taken into account that, at the phase transition, $\theta = \theta_c^{stat}$ where $\theta_c^{stat}$ is given by \eqref{eq: critical strength}. The above expression shows that the residue for a static phase transition is completely imaginary.
\\
In the main text, we have investigated the static phase transition arising in the homogeneous Kuramoto model characterised by a delta-distribution of frequency $\mu(h) = \delta(h)$. It is easy to see from \eqref{eq: static residue} that, in this case, the residue is independent of the parameter of the system and equals to $\kappa = \frac{i}{2}$.
\subsubsection{Residue for dynamic phase transitions}
We here investigate the residue for dynamic phase transitions characterised by the development of singularities of the susceptibility for a pair of opposite real frequencies $\omega_1 = - \omega_2 = \omega^* >0$.
\\
A similar argument of the previous section holds but it has to be modified to take into account both poles. We have that
\begin{equation}
    \tilde{\chi}_1(\omega) = \frac{Y_{11}(\omega)}{P(\omega)} = \frac{Y_{11}(\omega^*)}{P'(\omega^*)\left(\omega - \omega^* \right)} +  \frac{Y_{11}(-\omega^*)}{P'(-\omega^*)\left(\omega + \omega^* \right)} + \psi(\omega)
\end{equation}
where $\psi(\omega)$ is an analytic function in the upper complex $\omega$ plane. As before we define the residue $\kappa = \frac{Y_{11}(\omega)}{P'(\omega^*)}$ and we observe that, since the susceptibility, being the Fourier transform of a real function, has to satisfy the condition $\tilde{\chi}_i(-\omega) = \tilde{\chi}_i^*(\omega)$. This implies that $\frac{Y_{11}(-\omega^*)}{P'(-\omega^*)} = - \kappa^*$ 
so that the previous equation can be written as
\begin{equation}
\label{eq: proper equation for poles}
    \tilde{\chi}_1(\omega) = \frac{\kappa}{\omega-\omega^*} - \frac{\kappa^*}{\omega+\omega^*} + \psi(\omega),
\end{equation}
Similarly to the previous section $Y_{11}(\omega^*)=1/\theta$ and the residue is
\begin{equation}
\label{eq: residue dynamic transition formula}
    \kappa = \frac{1}{\theta P'(\omega^*)}
\end{equation}
As opposed to the static phase transition, the evaluation of the residue now requires a more careful study of $P'(\omega^*)$ where $\omega^* >0$. 
\\
We are here interested in the evaluation of the residue for the dynamic phase transition arising from the bimodal distribution $\mu_0$ discussed in the previous sections. We have seen that such transition is associated to a pole $\omega^* = \sqrt{h_0^2 - D^2}$ arising at the transition point $\theta_c^{dyn} = 4D$.
Evaluating \eqref{eq: derivative of P} for a bimodal distribution $\mu_0$ and separating real and imaginary part we obtain the following equations
\begin{subequations}
    \begin{align}
    P'(\omega)&=P_R'(\omega)+i P_I'(\omega) \\
P_R'(\omega^*) &= \frac{\theta D}{\left( D^2+h_0^2\right) }\left(C_I(\omega^*;h_0) + \omega^* C_{I}'(\omega^*;h_0) \right) \\
 P_I'(\omega^*) &= -\frac{\theta D}{\left( D^2 + h_0^2\right)}\left(C_R(\omega^*;h_0) + \omega^* C_R'(\omega^*;h_0) \right) =-\frac{\theta D}{\left( D^2 + h_0^2\right)} \omega^* C_R'(\omega^*;h_0) 
    \end{align}
\end{subequations}
where we have taken into account \eqref{eq: def of C} and in the last equality we have imposed condition \eqref{eq: condition bimodal}. We can evaluate the derivative of $C(\omega;h_0)$ from equations \eqref{eq: def of C},\eqref{eq: real part C} and \eqref{eq: imaginary part C} with similar arguments. We omit the details, as it is just algebra, but we mention that we make extensive use of the fact that $\omega^2_* := (\omega^*)^2 = h_0^2 + D^2$. The result is
\begin{subequations}
    \begin{align}
        C'(\omega^*;h_0) &= C_R'(\omega^*;h_0) + i C_I'(\omega^*;h_0) \\
        C_R'(\omega^*;h_0) &= \frac{\omega^*}{4D^3}\frac{D^2+h_0^2}{D^2+\omega_*^2} \\
        \label{eq: derivative of C imaginary bimodal}
        C_I'(\omega^*;h_0) &= - \frac{1}{4\left( D^2+\omega_*^2\right)}
    \end{align}
\end{subequations}
From the above equations it is immediate to evaluate 
\begin{equation}
    P_I'(\omega^*) = - \frac{\omega_*^2}{D^2+\omega_*^2} \frac{\theta}{4D^2} = - \frac{1}{D}\frac{\omega_*^2}{D^2+\omega_*^2}
\end{equation}
where we have used that $\theta = \theta_c^{dyn} = 4D$. We can then proceed to calculate $P'_R(\omega^*)$. Taking into account \eqref{eq: condition bimodal on imaginary part C} and \eqref{eq: derivative of C imaginary bimodal} we get
\begin{equation}
    P'_{R}(\omega^*) = \frac{\theta D}{ \left( D^2 + h_0^2\right)}\left(-\frac{\omega^*}{4D^2} - \frac{\omega^*}{4 \left( D^2 + \omega_*^2\right)}\right) = - \frac{\theta \omega^* \left(2D^2 + \omega_*^2 \right)}{4D\left(D^2 + h_0^2 \right)\left(D^2 + \omega_*^2 \right)} = - \frac{\omega_*}{D^2+\omega_*^2}
\end{equation}
where in the last equality we have used that, by the definition of $\omega^*$, $2D^2+\omega_*^2 = D^2+h_0^2$ and that $\theta = \theta_{c}^{dyn} = 4D$. As a result, we can write that 
\begin{equation}
    P'(\omega^*) = P_{R}'(\omega^*) + P_I'(\omega^*) = - \frac{\omega_*}{D^2+\omega_*^2}\left(1 + i \frac{\omega_*}{D} \right) 
\end{equation}
Since $\theta = \theta_c^{dyn} = 4D$, the residue \eqref{eq: residue dynamic transition formula} turns out to be for the dynamic phase transition associated to the bimodal frequency distribution
\begin{equation}
    \kappa = \kappa_R + i \kappa_I = - \frac{D}{4\omega^*} + \frac{i}{4}
\end{equation}
where $D$ satisfies the property $\theta = 4 D$ identifying the phase transition in the parameter space $(\theta,D)$. As opposed to the static transition, the residue for the dynamic phase transition is not completely imaginary but has also a real part that depends on both the strength of the noise and the critical frequency. Consequently, such singular behaviour is fundamentally different from the static transition as will be highlighted in the next section. 
\subsection{Response for finite $N$: singularities and resonances}
In order to numerically evaluate the response properties of the finite system we perform, following \cite{Marconi2008},  $n$ simulations of equations \eqref{eq appendix: NLFPE Kuramoto} where the initial conditions are sampled according to the unperturbed invariant measure $\rho_0 = \frac{1}{2 \pi}$ and where at time $t = 0$ we apply a perturbation proportional to a Dirac $\delta(t)$ function. We remark that this is a macroscopic perturbation where each agent is perturbed according to $F(x;h) \to F(x;h) + \varepsilon X(x)\delta(t)$, where $F(x;h)=h$ for the Kuramoto model. The average of the response $\langle \langle A_i \rangle \rangle = \langle \langle A_i \rangle \rangle_0 + \varepsilon \langle \langle A_i \rangle \rangle_1 = \varepsilon \langle \langle A_i \rangle \rangle_1$ over the $n$ simulations gives an estimate $\tilde{G}_i$ of the macroscopic response function defined since
\begin{equation}
\label{eq: time convolution}
  \langle \langle A_i \rangle \rangle_1(t) = \left(\tilde{G}_i \star T \right)(t) =  \left(\tilde{G}_i \star \delta \right)(t) = \tilde{G}_i(t)
\end{equation}
where we remark that $\langle\langle A_i \rangle\rangle_0 = 0$ for the Kuramoto model.  Convergence to the linear response regime has been assessed by performing simulations with different values of $\varepsilon$. The total susceptibility $\tilde{\chi}_i(\omega)$ of the system is obtained as 
\begin{equation}
    \tilde{\chi}_i (\omega) = \int_{- \infty}^{+\infty}\tilde{G}_i(t)e^{i \omega t}\mathrm{d}t
\end{equation}
For finite $N$ we observe a mollifying effect of the singular behaviour of the susceptibility \cite{ZagliLucariniPavliotis} where the poles $\omega^*$ gets slightly shifted towards the lower half of the complex plane, namely $\pm \omega^* \to \pm \omega^* - i \gamma(N)$ where $\gamma(N) \to 0$ as $N \to + \infty$. Singularities of the susceptibility in the thermodynamic limit correspond to resonances for finite $N$ characterised by 
\begin{equation}
\label{eq: resonances}
    \tilde{\chi}_1(\omega) =  \frac{\kappa}{\omega-\omega^*+ i \gamma(N)} +\psi(\omega) = \frac{\kappa}{\omega-\omega^*+ i \gamma(N)} - \frac{\kappa^*}{\omega+\omega^*+ i \gamma(N)} + \psi '(\omega)
\end{equation}
The resonant behaviour of the susceptibility becomes singular in the thermodynamic limit as, for $N \to + \infty$,
\begin{equation}
\label{eq: asymptotic behaviour}
    \lim_{N \to \infty} \tilde{\chi}_N(\omega) = - i \pi \kappa \delta(\omega-\omega^*) + \kappa \mathcal{P} \left( \frac{1}{\omega-\omega^*} \right) + \psi(\omega).
\end{equation}
where $\mathcal{P}(\frac{1}{\omega})$ represents the Principal Value distribution. For the static transition, the residue is completely imaginary so that $\mathbf{Re}\tilde{\chi}_i(\omega)$ provides the $\delta-$diverging behaviour whereas the imaginary part yields the Principal Value behaviour. In particular, the function $\omega \mathbf{Im}\tilde{\chi}_i(\omega)$ is constant and equals to $|\kappa|$ in a neighborhood of $\omega^*=0$. A visual inspection of Panel $(a)$ (bottom inset) of Figure $1$ in the main text clearly shows that numerical experiments agree with the theoretical prediction $|\kappa| = \frac{1}{2}$ for the homogeneous case $\mu(h) = \delta(h)$.
\\
As for the dynamic transition, the residue has a real part too and the above identification of the two different behaviours of the susceptibility is not as straightforward. More importantly, the non vanishing real part of the residue has strong consequences on the response properties of the system, with the development of a spontaneous phase shift of the response with respect to the forcing. In order to clarify this feature, we evaluate \eqref{eq: resonances} at the critical frequency $\omega=\omega^*$ and neglect non diverging terms in $\gamma(N)$
\begin{equation}
    \tilde{\chi}_1(\omega^*) = \frac{\kappa}{i \gamma(N)} = \frac{\kappa_I}{\gamma(N)} -  i \frac{\kappa_R}{\gamma(N)}.
\end{equation}
We now consider a pure sinusoidal time dependent forcing $T(t;\hat \omega) = \sin(\hat \omega t)$ and evaluate the response  $\langle \langle A_1 \rangle \rangle_1$ through the response formula 
\begin{equation}
\label{eq: basic linear response}
    \langle \langle A_1 \rangle \rangle_1(\omega) = \tilde{\chi}_1(\omega)T(\omega) 
\end{equation}
Given that $T(\omega;\hat \omega)=- i \pi \left( \delta(\omega+\hat \omega) - \delta(\omega - \hat \omega ) \right)$, by taking the anti-Fourier Transform of \eqref{eq: basic linear response} one obtains the response in the time domain
\begin{align}
\label{eq: response to sinusoidal}
    \langle \langle A_1 \rangle \rangle_1(t;\hat \omega) &= - \mathbf{Im}\left( e^{-i\hat \omega t} \tilde{\chi}(\hat \omega)\right) = - \mathbf{Im} \tilde{\chi}(\hat \omega) \cos(\hat \omega t) + \mathbf{Re} \tilde{\chi}(\hat \omega) \sin(\hat \omega t)=  \\
    &= |\tilde{\chi}(\hat \omega)|\sin\left( \hat \omega t+ \arctan\left( \frac{\mathbf{Re}\tilde{\chi}(\hat \omega)}{\mathbf{Im}\tilde{\chi}( \hat \omega)}\right) - \frac{\pi}{2} \right)
\end{align}
The response to perturbations at the critical frequency $\omega^*$ is then
\begin{equation}
\label{eq: critical response at critical frequency}
    \langle \langle A_1 \rangle \rangle_1(t;\omega^*) = \frac{1}{\gamma(N)} \left( \kappa_R \cos(\omega^* t) + \kappa_I \sin(\omega^* t) \right) = \frac{|\kappa|}{\gamma(N)}\sin\left( \omega^*t + \varphi ' \right)
\end{equation}
where  $\varphi' = \varphi - \frac{\pi}{2}$ and $\varphi$ is related to the ratio of the imaginary and real part of the residue as 
\begin{equation}
\label{eq: shift}
\varphi = \arctan\left(- \frac{\kappa_I}{\kappa_R} \right) = \arctan\left(\frac{\omega^*}{D} \right)  
\end{equation}
Firstly, as expected, the amplitude of the oscillation diverges in the thermodynamic limit as $\gamma(N) \to 0$. Secondly, given that $\kappa_R \neq 0$, the response spontaneously develop a phase shift $\varphi ' \neq 0$ with respect to the external sinusoidal forcing. 
We have investigated the development of the phase shift by comparing the numerical response $\langle \langle A_1 \rangle \rangle_1(t)$ obtained as the convolution product \eqref{eq: time convolution} and the exact theoretical result \eqref{eq: critical response at critical frequency}. A quantitative estimation of the phase shift is obtained by looking at the correlation function between the forcing and the response. The values at which the correlation function has its maxima correspond to the phase shift. A visual inspection of figure \ref{fig:phase shift critical} clearly indicates a very good agreement between numerical results and theory. It is also interesting to investigate the response of the system to a very fast driving signal. This regime can be investigated by looking at the high frequency behaviour of the critical susceptibility of the system. We expect that $\mathbf{Re}\tilde{\chi}_i(\omega) \approx 0$ for $\omega \to \infty$ so that, from \eqref{eq: response to sinusoidal}, the response will be in quadrature with the forcing $\langle \langle A_1 \rangle \rangle (t) \propto \cos(\omega t)$, as clearly shown in the right column of figure \ref{fig:phase shift critical} (here we have chosen a frequency $\hat \omega = 2 \omega^*$).  
\subsubsection{Parameters for the numerical experiments}
We here report the critical and off-critical parameters for the numerical experiments. 
\begin{itemize}
    \item Static Transition $\mu(h) = \delta(h)$: the coupling strength is fixed$\theta = 1$. The value of the noise strength at the phase transition is $\sigma_c = 1$, whereas away from the transition is $\sigma = 1.1$. 
    \item Dynamic Transition $\mu_{h_0}(h) = \frac{1}{2}\left( \delta(h-h_0) + \delta(h+h_0)\right)$: $h_0= 2$, $\theta = 2$ are fixed. The value of the noise strength at the transition is $\sigma_c = 1$, away from the transition is $\sigma = 1.4$. As a result, the critical frequency is $\omega^* = \sqrt{h_0^2 - D^2} \approx 1.93$ and the phase shift $\varphi ' = \varphi - \frac{\pi}{2}  \approx -0.25$, see equation \eqref{eq: shift}.
\end{itemize}

\end{document}